\documentclass[compsoc,conference,a4paper,10pt,times]{IEEEtran}
\usepackage{mathptmx}  % Postscript fonts for math mode
\usepackage{amsmath}
\usepackage{amsfonts}
\usepackage{amssymb}
\usepackage{graphicx}

\title{Decentralized Lightweight Detection of Eclipse Attacks on Bitcoin Clients}
\author{Bithin Alangot, Daniel Reijsbergen, Sarad Venugopalan and Pawel Szalachowski \\
    \small{Singapore University of Technology and Design, Singapore}\\
}

\usepackage{tikz}
\usepackage[english]{babel}
\usepackage[colorlinks=true,urlcolor=black]{hyperref}
\addto\extrasenglish{%
}

\usepackage{multirow}
\usepackage{booktabs}
\usepackage{caption}
\usepackage{framed}
\usepackage{listings}
\usepackage[ruled,linesnumbered]{algorithm2e}
\usepackage{subcaption}
\usepackage{xcolor, colortbl}    % Used to highlight comments
\usepackage{soul}
\usepackage{pifont}
\usepackage{balance}

\date{}

\def\BibTeX{{\rm B\kern-.05em{\sc i\kern-.025em b}\kern-.08em
    T\kern-.1667em\lower.7ex\hbox{E}\kern-.125emX}}
\begin{document}

\maketitle

\begin{abstract}
    Clients of permissionless blockchain systems, like Bitcoin, rely on an
    underlying peer-to-peer network to send and receive transactions.  It is
    critical that a client is connected to at least one honest peer, as
    otherwise the client can be convinced to accept a maliciously forked view of
    the blockchain.  In such an \textit{eclipse attack}, the client is unable to
    reliably distinguish the canonical view of the blockchain from the view
    provided by the attacker. The consequences of this can be
    catastrophic if the client makes business decisions based on a distorted
    view of the blockchain transactions. 
    
    In this paper, we investigate the design space and propose two approaches
    for Bitcoin clients to detect whether an eclipse attack against them is
    ongoing. Each approach chooses a different trade-off between average attack
    detection time and network load. The first scheme is based on the detection
    of suspicious block timestamps. The second scheme allows blockchain clients
    to utilize their natural connections to the Internet (i.e., standard web
    activity) to gossip about their blockchain views with contacted servers and
    their other clients. Our proposals improve upon previously proposed eclipse
    attack countermeasures without introducing any dedicated infrastructure or
    changes to the Bitcoin protocol and network, and we discuss an
    implementation. We demonstrate the effectiveness of the gossip-based schemes
    through rigorous analysis using original Internet traffic traces and
    real-world deployment. The results indicate that our protocol incurs a
    negligible overhead and detects eclipse attacks rapidly with high
    probability, and is well-suited for practical deployment.
\end{abstract}


\section{Introduction}
\label{sec:intro}

The invention of \emph{blockchains} as a means of providing an immutable,
trustless, and de-centralized ledger promises to revolutionize monetary
transfers alongside other applications. It has enabled cryptocurrencies such as
Bitcoin~\cite{nakamoto2008bitcoin} to grow without the need for a central
authority, and due to its transparency has applicability in a broad range of
fields including secure logging infrastructures~\cite{tomescu2017catena},
distributed timestamping~\cite{opentimestamp}, and micropayment
channels~\cite{decker2015fast}. Any user can join the Bitcoin network and view
or add transactions to the distributed ledger. Since the Bitcoin network is an
overlay network~\cite{wang2018survey} on top of a public infrastructure (the
Internet), a number of attacks on the underlying network are inherited by the
Bitcoin network. For example, attacks on DNS servers arise from a lack of proper
authentication, or misaligned trust between entities in the delegated DNS
hierarchy. Most Bitcoin clients, including the popular \textit{bitcoinj}, depend
on DNS seeders~\cite{bitcoinj} to resolve the symbolic names of peers to their
respective IP addresses. A recent survey reveal that 95\% of cryptocurrencies
employ DNS seeding or IP hard-coding which are censorship prone
techniques~\cite{loe2019you}. If the DNS server used for name resolution by the
client is compromised or malicious, then the DNS cache can be
poisoned~\cite{Son2010DNS} to resolve the domain name to an attacker-controlled
IP address. Similar attacks can be launched via insecure Internet core protocols
like BGP. As discussed in the literature~\cite{heilman2015eclipse,
apostolaki2017hijacking, marcus2018low, ekparinya2018impact,
gervais2015tampering, tran2020stealthier}, such network-level vulnerabilities
make Bitcoin clients vulnerable to \emph{eclipse attacks}, where an attacked
client is connected only to attacker-controlled peers.  

In this paper, we propose \emph{two} new protocols to help Bitcoin clients
detect whether they are being eclipsed. The first protocol detects eclipse
attacks by using suspicious block timestamps -- i.e., if the time between newly
created blocks is too high, then this indicates that the network has been
partitioned. 

This protocol can be executed by any client in isolation, however, our analysis
shows that it takes around 2-3 hours for a client to be relatively sure that it
is under attack. To reduce the average attack detection times, we also propose a
\textit{ubiquitous gossip protocol} in which the client piggybacks gossip
messages onto its connections to protocol-running servers. This protocol does not
require any changes to Bitcoin or its peer-to-peer network. It also requires
minimal support from web servers to communicate a small number of Bitcoin block
headers. A server can be any host on the Internet that participates in the
gossip protocol as a service to the public, and which contributes a small amount
of data storage (in the order of few kilobytes) to store the gossip messages. A
gossip message consists of a set of Bitcoin block headers. The server receives
these gossip messages as different clients connect to it, and the server
maintains the strongest view of the blockchain it has seen yet. This view is
then passed to the connecting clients in exchange. The gossip protocol can
function in a \textit{passive} or \textit{active mode}. In the passive mode,
block headers are gossiped via inclusion in the HTTP(S) traffic when a client
makes a standard web connection with the server. In the active mode, the client
initiates connections with known protocol-running servers for the sole purpose
of gossiping (to update its view of the blockchain) before it conducts a wallet
balance check. We demonstrate the efficiency of our passive mode attack
detection via a thorough analysis of real-world Internet traffic traces and show
that, on average, the attack detection happens in less than an hour, depending
on how many commonly-visited servers participate in the protocol. This time
duration is typically sufficient since a block confirmation takes approximately
one hour (due to the 6 block confirmation rule~\cite{bitcoinconfirmations},
while in reality more confirmations may be required due to mining pool
centralization). 

However, in active mode the detection is almost instantaneous, although
additional connections with protocol-running servers are necessary. These
servers are identified through passive-mode gossiping.
The requirements and attack detection time for the proposed protocols are
summarized in~\autoref{tab:comparison}. 

Various countermeasures have been proposed in the literature to detect eclipse
attacks -- these are discussed in~\autoref{sec:related}. Our goal is to improve
on those approaches by providing a practical solution for Bitcoin clients to
detect eclipse attacks. The main contributions of this paper are to
$\romannumeral 1$) propose the first (up to our best knowledge) lightweight  and
ubiquitous gossip protocol that can detect eclipse attacks on Bitcoin clients,
$\romannumeral 2$) present a fully passive eclipse attack detection protocol
based on ``suspicious'' block timestamps which does not incur additional
overhead or server participation, $\romannumeral 3$) propose an
implementation of the gossip protocols, and $\romannumeral 4$)
thoroughly analyze the effectiveness of our protocols using
real-world network traffic traces and blockchain data.

The rest of the paper is organized as follows.~\autoref{sec:background} provides
 background information on Bitcoin and eclipse attacks.~\autoref{sec:related} reviews
the related work. ~\autoref{sec:overview} provides an overview of the
requirements and system model for the proposed the eclipse detection
protocols.~\autoref{sec:baseline} presents the timestamp-based
protocol.~\autoref{sec:gossipoverview} presents our gossip-based protocol and
analyzes the real Internet traffic trace to demonstrate its
efficiency.~\autoref{sec:discussion} discusses real-world deployment and
privacy concerns and~\autoref{sec:implementation} details the implementation and
evaluation.~\autoref{sec:conclusions} concludes the paper.

\section{Bitcoin and Eclipse Attacks}
\label{sec:background}
\begin{table*}[h]
	\centering
	
	\begin{tabular}[t!] {c c c c c c}
		\toprule
		\multicolumn{2}{c}{} & {\textbf{Infrastructure Support Required}} & {\textbf{Network Load}} & 
		{\textbf{Attack Detection Time}} & {\textbf{Refer to}} \\
		\toprule
	 
		\textbf{Timestamp} & &  None & None  &  3 hours & \autoref{sec:baseline} \\\midrule
		   
		\multirow{2}{*}{\textbf{Gossip}} & passive mode & \hfil Servers & Natural browsing traffic & \hfil $>$1 hour & \multirow{2}{*}{\autoref{sec:gossipoverview}} \\
	
		 & active mode & \hfil Servers & Additional connections & \hfil Immediate &  \\
		\bottomrule
		
	\end{tabular}
	\caption{Comparison of the proposed eclipse attack detection protocols.}
	\label{tab:comparison}
\end{table*}

\subsection{Bitcoin}
Bitcoin is a distributed, peer-to-peer electronic payment system that enables
Internet-based payments without going through a centralized and trusted entity
like a financial institution~\cite{nakamoto2008bitcoin}. A distributed copy of
the electronic transaction ledger is stored by multiple peers of the Bitcoin
network.  Transactions are grouped into blocks, and the blocks are
cryptographically linked forming a chain (called a blockchain).  Bitcoin uses
proof-of-work (PoW) for its consensus mechanism. A number of peers
simultaneously attempt to solve a puzzle by finding a pre-image of a
cryptographic hash that satisfies the condition of the puzzle. This process is
called mining.  A peer that solves the puzzle broadcasts its block to other
peers in the network.  Peers are incentivized to create blocks as they receive a
block reward and transaction fees (in Bitcoin's native cryptocurrency called
\emph{bitcoin}).  Two concurrent blocks can be mined and announced
simultaneously, and in such a case peers accept the block they receive first.
This process of disagreement between peers is called forking.  Forking in
Bitcoin is resolved by all honest peers agreeing to follow the strongest chain
rule, where a chain with the most PoW aggregated is considered to be the current
one. Bitcoin requires that a majority of computational power (to mine blocks)
belongs to honest parties, so that they can resolve forks in their favor and
hence prevent double-spending attacks~\cite{Chohan2017}. 

In Bitcoin, a block consists of two parts: the header~\cite{block_headers} and
the list of transactions~\cite{transaction}. The Bitcoin header aggregates the
transactions and contains metadata -- in particular, the following fields:
\textit{timestamp}, which is the (approximate) time at which the block was
created, \textit{prevHash}, which is the hash pointing to the block preceding
the current block (this effectively creates a chain and provides an ordering to
the blocks), and \textit{nBits}, which encodes the block's difficulty
requirement (i.e., the amount of PoW effort required to find a solution hash to
the mining puzzle).  The difficulty is dynamically set every 2016 blocks in a
way to adjust the average block creation time to 10 minutes.  At this rate, 2016
blocks would be created in exactly 2 weeks. If it took more than 2 weeks to
generate the 2016 blocks, the difficulty is reduced by adjusting the value of
\textit{nBits}. A time shorter than 2 weeks results in a difficulty increase,
and a longer time in a decrease.

Peers in the Bitcoin network can, depending on their resource constraints, be
categorized as mining nodes, full nodes, or simplified payment verification
(SPV) nodes. Throughout the discussion, SPV nodes are also referred to as
lightweight clients, light clients or clients. A mining node stores and verifies
every block in the blockchain and competes to mine new blocks. A full node is
similar to a mining node except that it does not work to mine blocks. An SPV
client is a node that stores the block headers alone and verifies that a) each
block header points to the previous block header b) and each block header was
generated with the PoW required. Bitcoin aggregates transactions within blocks
in a way that allows SPV clients (who only store block headers) to verify that
any included transaction is part of the blockchain.  To check whether a
transaction has made it onto the blockchain, an SPV client makes an API call to
a full node to request a proof. 

\subsection{Eclipse Attacks on Bitcoin}
To handle the communication, Bitcoin introduces a peer-to-peer Internet overlay
network.  When a  peer attempts to connect to the Bitcoin network, it first
finds initial (seed) peers via a DNS lookup to pre-defined domain names.  A
successful DNS lookup allows the peer to contact seed peers, which in turn
return lists of their known peers.  The peer connects to these peers and can
hence start using the protocol.  The security of a Bitcoin client relies on its
ability to connect to honest peers in the Bitcoin network. Being connected to an
attacker-controlled Bitcoin network undermines the ability of the client to view
or transact on the honest ledger, which can result in financial losses. The
client may assume that it received payment in exchange for goods and services,
only to realize later that the transaction was registered only on the
attacker-controlled blockchain and not the canonical blockchain. Detecting such
attacks is hence critical for the security of Bitcoin clients.

Unfortunately, as the Bitcoin network relies on the Internet, it inherits all
its security drawbacks.  Most prominently, in eclipse
attacks~\cite{heilman2015eclipse,Heilman2015presentation} an adversary manages
to hijack all connections from an attacked client to other peers in the network.
The client's view of the network and information  dissemination is fully under
the control of the attacker. The attacker can provide a malicious view of the
Bitcoin blockchain to the client, which may include transactions of bitcoins
that have already been spent on the unseen part of the blockchain's canonical
branch (this is called a \textit{double-spend} attack).  A Bitcoin lightweight
client may be vulnerable to an eclipse attack through DNS cache
poisoning~\cite{Son2010DNS} involving the name resolution of Bitcoin
seeders~\cite{bitcoingithubrepo01152019,bootstrapservers}, or routing attacks
where the Bitcoin network is partitioned~\cite{apostolaki2017hijacking} and the
peer eclipsed.

\section{Related work}
\label{sec:related}
In this section, we investigate network-level attacks on blockchain systems and
their countermeasures from the existing literature. Heilman et
al.~\cite{heilman2015eclipse} studied and presented the feasibility of an
eclipse attack on the Bitcoin network. They showed that it is viable for a
powerful adversary who controls a large number of public IPs to monopolize all
peer connections to a victim Bitcoin client and consequently present a malicious
Bitcoin blockchain view. The ability of an adversary to use this vulnerability
to conduct selfish mining and double-spend attacks was further discussed by
Gervais et al.~\cite{gervais2016security} and Nayak et
al.~\cite{nayak2016stubborn}. Another class of serious attacks is connected to
the BGP protocol which is one of the core Internet protocols. Apostolaki et
al.~\cite{apostolaki2017hijacking} demonstrated that the Bitcoin protocol is
vulnerable to BGP routing attacks where an attacker controlling a small number
of BGP prefixes can partition the Bitcoin network by announcing malicious BGP
messages.  To prevent such an attack, the SABRE framework was
proposed~\cite{apostolaki2018sabre}. It is a secure relay network which helps to
protect against BGP routing attacks by enabling the Bitcoin clients to connect
to relay nodes hosted at safe autonomous systems. To prevent (D)DoS attacks on
relay nodes this architecture requires high-performant programmable network
switches. Tran et al. demonstrated that Bitcoin clients are vulnerable to the
Erebus attack~\cite{tran2020stealthier}, which is a data plane attack (in
contrast to the BGP routing attack, which is a control plane attack) and hence
requires no routing manipulation. This makes it much more stealthy than previous
attacks. Here, an attacker who is able to control ASes that can intercept
traffic to a specific set of public IPs which need not be Bitcoin client
addresses can execute Erebus attacks. However, in this study (as in all studies
above) a successful attacker needs to be powerful (controlling a large number of
public IPs or having access to BGP routers). In contrast, W{\"u}st and
Gervais~\cite{wust2016ethereum} as well as Marcus et al.~\cite{marcus2018low}
showed the feasibility of an eclipse attack on the Ethereum blockchain by
exploiting vulnerabilities in its peer-to-peer protocol where an attacker needs
only a small number of machines with public IPs to compromise a victim.
Recently, Loe and Quaglia~\cite{loe2019you} presented a survey which shows that
95\% of existing cryptocurrenies are using censorship prone technique to
bootstrapping i.e to identify peers in the network. The reason for this is
attributed to code reuse of the five major cryptocurrenies. Also, the survey
highlight the fact that 32\% of cryptocurrenies rely on a single DNS provider
for their DNS seeds leading to single point of failure. Finally, they analyze
censorship resilient techniques which was found to be very inefficient with high
latency overhead and none of the cryptocurrenies were able connect using this
technique.

Although we are not aware of any work similar to ours, the detection and prevention of
attacks on blockchain light clients is an active research topic. The
popular Bitcoin light clients add hard-coded
\textit{checkpoint} block headers~\cite{bitcoinjcheckpoint, electrumcheckpoint} 
into their code bases. This helps to prevent malicious miners from reorganizing
large parts of blockchain to produce a weaker view and present it to the
light client. The bitcoinj light client developers have mentioned a
proposal to detect eclipse attacks by analyzing the block arrival
rate~\cite{bitcoinj}, but there is no known implementation of such a feature
yet. In~\autoref{sec:baseline}, we provide a detailed study on the effectiveness
of such a feature in protecting light clients against eclipse attacks. 

The \textit{fraud proofs} by Mustafa et al.~\cite{al2018fraud} help light
clients to identify invalid blocks and reject them. The proposed data
availability proofs support the light clients to gossip small chunks of
information about the block for which it received a fraud proof. Therefore, the
network which consists of light clients and full nodes can rebuild the complete
block information to validate the proof. However, this approach requires
significant modifications to the existing protocols and needs the support of a
threshold number of honest light clients.  The protocol is able to detect
incorrect blockchain blocks, however, in contrast to our scheme fraud proofs do
not detect eclipse attacks if the blocks have a valid content.

One research direction is to make light clients even lighter such that they do
not have to process entire chains.  Non-Interactive Proofs of Proof-of-Work
(NIPoPoW)~\cite{kiayias2017non} and FlyClient~\cite{luuflyclient} help encode
blockchains such that their total PoW is expressed in a concise manner.  For
every chain, light clients only need to download a small proof from full nodes
to make sure that their view is stronger than all known alternatives.
Unfortunately, these schemes involve significant modifications to the Bitcoin
protocol, e.g., Bitcoin header modification. Also, the security of NIPoPoWs is
guaranteed only under certain parameter settings and in some cases it involves
multiple round-trip communications between the light clients and full nodes,
which increases the overhead at the light client. Although many of these issues
were addressed in FlyClient, their protocol modifications form an obstacle to
adaptability. Similar to the previous work, neither NIPoPoWs nor FlyClient can
detect eclipse attacks by themselves -- however, if deployed they could minimize
the overheads introduced by our system.

Conceptually, the most related work to ours is in the context of monitoring the
consistency of append-only centralized log servers as presented by Chuat et
al.~\cite{chuat2015efficient}.  They propose a protocol for monitoring the
consistency of certificate logs~\cite{rfc6962}, where web clients exchange
signed log statements via their HTTP(S) connections. This process helps to find
potential inconsistencies  in log statements, hence proving malice.  An
alternative approach for the same problem was proposed by Nordberg et
al.~\cite{nordberg2015gossiping}, where web clients implement a feedback
mechanism to inform a domain about observed log statements for the domain
certificates.  Due to the different setting, these schemes are not applicable in
our scenario.

\section{Solution Overview}
\label{sec:overview}
The objective of this work is to provide lightweight methods for a
Bitcoin light client to detect that it is being subjected to an eclipse attack.

Our first observation is that if a Bitcoin light client is under an
eclipse attack, it is then impossible for it to detect the attack via the
Bitcoin network itself (as its view is controlled by the adversary).  Therefore,
to facilitate detection of the attack, an external infrastructure is necessary.
However, deploying a new dedicated infrastructure is a challenging task in
practice.  Therefore, the protocol would ideally be implemented on top of a
currently existing infrastructure. 

The second observation is that although the attacked light client cannot
itself learn that its blockchain view is malicious, the attack is trivially
detectable if any stronger concurrent view of the blockchain is available to the
affected client. Permissionless blockchains, like Bitcoin, do not
introduce any trusted entities that could assert which blockchain view is
canonical. Instead they follow the strongest chain rule, so given two
conflicting views of different strengths it is trivial to decide which is the
canonical one.  Therefore, the deployed detection infrastructure can be
implemented as a medium for exchanging blockchain views between protocol
participants. Every participant can simply compare its local view with the
obtained one, detect any potential attack, and save the strongest view as the
current one.

The main idea behind our gossip-based protocols is based on the observation that
users of Bitcoin light clients conduct standard network connections, like web
browsing, messaging, email, etc, even when their client is under an eclipse
attack. Therefore, if they were to be able to exchange their blockchain views
with contacted Internet servers who would store only the strongest seen view,
then that could be the base for a detection system. The passive gossip scheme
only uses natural traffic for attack detection. By contrast, the active gossip
scheme uses natural traffic to learn of protocol-following nodes, and purposely
initiates connections with a subset of them during periods of interest, i.e.,
when the user checks her wallet balance via the client.  The timestamp-based
approach uses its own principles, which will be discussed in
\autoref{sec:baseline}. In the remainder of this section we present our system
model and the requirements of our detection systems.

\subsection{System Model}
\label{sssec:SystemModel}
The main agents in our system are as follows:

\begin{itemize}
    \item \textbf{Server (S)} is an Internet server that provides services
        accessible to the public and supports our protocol. Each server
        builds and stores its view of the blockchain from the block headers
        supplied to it by the clients when they establish a connection
        with the server. Each client sends a partial consecutive view of
        its block headers to the server when it connects to use its service.
	
    \item \textbf{Lightweight Client (LC)} (or just a client) is an SPV node of the
        Bitcoin network. Each LC obtains its view of the blockchain from the
        Bitcoin network that it is connected to. An LC may be under an eclipse
        attack, in which case it receives a malicious view of the blockchain
        controlled by an attacker.  We assume that LCs beside Bitcoin software
        run other programs (e.g., a web browser) and conduct user-driven
        Internet connections (e.g., browsing or messaging).  In this work we
        focus on light clients as they are the most popular and convenient way
        for regular users to interact with the Bitcoin network. However, our
        protocol can be also run by other Bitcoin clients, e.g., full nodes.
		
    \item \textbf{Attacker} is a malicious entity able to launch an eclipse
        attack on Bitcoin clients. The adversary may control (either directly or
        indirectly) enough mining power to construct an inferior branch prefixed
        with the canonical view of the blockchain in an attempt to masquerade as
        the canonical Bitcoin blockchain.  The objective of the attacker is to
        partition the view of the Bitcoin network and provide a malicious view
        of the blockchain to the LC unnoticed. We assume that the adversary is
        not able to generate a stronger blockchain than the honest participants
        (or else the attacker could perform a so-called 51\% attack without the
        need to eclipse clients).

\end{itemize}

\subsection{Requirements}
\label{sssec:Requirements}
In order to realize a successful detection framework, we define the following
requirements:

\begin{LaTeXdescription}
    \item[Effectiveness:] participants of the detection framework should be able
        to detect ongoing eclipse attacks with high probability and speed. As in
        Bitcoin, the suggested transaction confirmation time is about one hour
        (i.e., six new-coming blocks after the transaction was appended), we aim
        for a similar time frame for attack detection.

    \item[Low overheads:] protocol-introduced overheads should be negligible. In
        particular, the scheme should not require high CPU, memory, and storage
        utilization, and it should not introduce high bandwidth overheads or
        latency inflation into the existing client-server communications.  The
        protocol should support a variety of light clients, including
        resource-constrained nodes like IoT devices.
   
    \item[Deployability:] the protocol should be deployable without any
        dedicated network infrastructure. Thus, the protocol should not incur
        any significant investments or setup efforts. It should use the existing
        infrastructure with minimal or no changes to the applications deployed
        on it.

   \item[Backward compatibility:] the scheme should not require any changes to
       the Bitcoin protocol or its network. As witnessed from past developments
       and deployments, the Bitcoin community is reluctant to introduce changes
       to the protocol, so if such a change is required then it could undermine
       the deployment of the attack detection framework.
    
\end{LaTeXdescription}

\section{The Timestamp-Based Protocol}
\label{sec:baseline}
The first eclipse attack detection protocol that we present is fully passive and
requires only the block timestamps, which are part of the block headers and
therefore known to the light clients by default. Of the three presented
approaches, this one has the slowest average detection time, but it is the
easiest to implement as it does not depend on protocol-running servers.

During an eclipse attack, an attacker can convince a lightweight client to
accept an inferior branch of the blockchain. However, to \emph{build} such a
branch the attacker will still need to control, either directly or indirectly, a
considerable amount of mining power. Since Bitcoin's mining difficulty changes
infrequently, the difficulty can be assumed to remain constant for the duration
of the attack. Since the attacker cannot create blocks at the same frequency as
the whole network, a sudden increase in the block creation times is likely. The
lightweight client \emph{bitcoinj} has indicated that in the future, it may implement a
``red alert'' mode based on the block creation times~\cite{bitcoinj}. To the
best of our knowledge, none of the lightweight clients have implemented
such a feature yet, so we present our timestamp-based protocol in this section.

\subsection{Block Timestamp Model}

Alerts in the timestamp-based approach are triggered by abnormally long block creation
times. Since block creation times are random by nature, we need a probabilistic
model. The time between block creations follows an exponential
distribution~\cite{decker2013information}, so the time needed to create $k$
blocks in a row follows the \emph{Erlang distribution} with shape parameter $k$
and a mean of $k \cdot 10$ minutes. However, day-to-day changes in the total
network hash rate mean that these assumptions are not always valid. We
investigate this using historical block timestamp observations and hash rate
estimates that were obtained via the data API of  {\tt
blockchain.com}.\footnote{\url{https://www.blockchain.com/api/blockchain_api}}
In particular, we collected timestamps for the blocks between height 506000 and
560013, which were mined on Jan.\ 25, 2018 and Jan.\ 25, 2019 respectively, and
which have the following hashes:

\noindent\resizebox{0.48\textwidth}{!}{\footnotesize{\texttt{0000000000000000000d69a840ca2ad3560d596ccc4d2c26e7e56f4b5d18ec4e}}}\newline
\resizebox{0.48\textwidth}{!}{\footnotesize{\texttt{0000000000000000003cd1f6db7b2e2009e975e3baac7fc8d1c1e53f8025b8d8}}}

\noindent For the hash rates we collected the data underlying the hash rate
estimate chart\footnote{\url{https://www.blockchain.com/charts/hash-rate}} from
the 2-year period starting from 23 Jan.\ 2017.

As we can see in~\autoref{fig:exponential_qq}, the hash rate has changed
considerably over time --- it tripled between Jan.\ 2018 and its peak around 18
Sept.\ 2018, and fell by roughly 1\% per day in the 32 days from 6 Nov.\ to 7
Dec. The observed average block creation time between 25 Jan.\ 2018 and 25 Jan.\
2019 was not 600 seconds, but 581, which reflects the fact for most of this
one-year period, the hash rate was increasing and the difficulty therefore often
too low. However, higher block creation times were observed when the hash rate
was dropping. In~\autoref{fig:exponential_qq}, the observed timestamp
differences are compared to an exponential distribution with a mean of 581
seconds: the fit is generally good, but more low and high extremes appear than
one would expect from an exponential distribution. The assumptions underlying
our calculations will be loosened to reflect this.

\begin{figure}[!t]
    \centering
\includegraphics[width=0.48\textwidth]{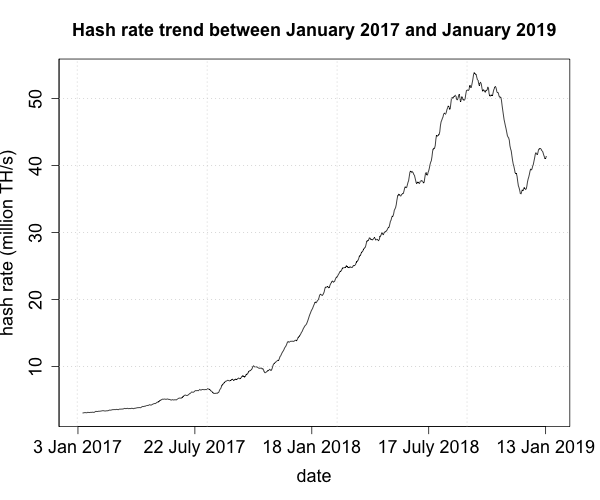}
\caption{Bitcoin hash rate between January 2017 and January 2019, smoothed using
a 14-day moving average filter (7 days on both sides).}
\label{fig:hashrate2years}
\vspace{-2mm}
\end{figure}

\begin{figure}[!t]
    \centering
\includegraphics[width=0.48\textwidth]{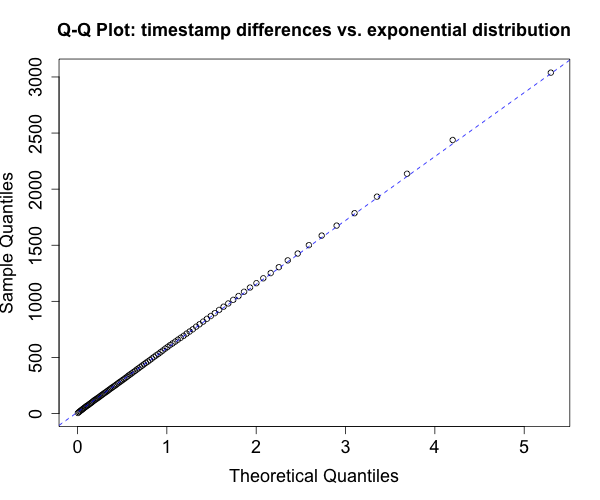}
\caption{Comparison between the observed timestamp differences from 25 January
2018 to 25 January 2019 (with negative values removed) and the exponential
distribution with the same mean as observed (581 seconds), via a Q-Q plot.
Better alignment between the circles and the dashed line means a close fit.
Despite the seemingly good fit, many uncharacteristically low and high values
are observed, and the Kolmogorov-Smirnov test rejects the null hypothesis that
the sample is drawn from the exponential distribution with a $p$-value of $4.565
\cdot 10^{-7}$. This is most likely due to considerable variations in the hash
rate during this period, as observed in~\autoref{fig:hashrate2years}. Moreover,
the sample size is very large (namely $>50,000$), which means that even minor
deviations from the exponential distribution will result in a low $p$-value.}
\label{fig:exponential_qq}
\vspace{-3mm}
\end{figure}

Since the largest observed drop in the hash rate trend since early 2017 has been
$1\%$ per day, and the Bitcoin difficulty resets every two weeks (or 16 days if
we account for the dropping hash rate), we will assume that at any time the
total network hash rate is at most 14-16\% lower than during the last difficulty
rescale. Hence, we will assume in the following that the time needed to create
$k$ blocks can be conservatively assumed to be Erlang-distributed with shape $k$
and a mean of $k \cdot {12}$ minutes (instead of $k \cdot 10$ minutes). 

\setlength{\tabcolsep}{3pt}
\begin{table*}[!ht]
\centering
\small

\begin{tabular}{ccccccccc|c|c} 
 & \multicolumn{10}{c}{number of observed blocks $n$} \\
 $t$ (mins) & 0 & 1 & 2 & 3 & 4 & 5 & 6 & 7 & 12 & 18\\ \hline
20 & \cellcolor{ green!30 }  $ 1.9 \cdot 10^{ -1 }$ & \cellcolor{ green!30 }  $
5.0 \cdot 10^{ -1 }$ & \cellcolor{ green!30 }  $ 7.7 \cdot 10^{ -1 }$ &
\cellcolor{ green!30 }  $ 9.1 \cdot 10^{ -1 }$ & \cellcolor{ green!30 }  $ 9.7
\cdot 10^{ -1 }$ & \cellcolor{ green!30 }  $ 9.9 \cdot 10^{ -1 }$ & \cellcolor{
green!30 }  $\approx 1$ & \cellcolor{ green!30 }  $\approx 1$ & \cellcolor{
green!30 }  $\approx 1$ & \cellcolor{ green!30 }  $\approx 1$ \\
40 & \cellcolor{ green!30 }  $ 3.6 \cdot 10^{ -2 }$ & \cellcolor{ green!30 }  $
1.5 \cdot 10^{ -1 }$ & \cellcolor{ green!30 }  $ 3.5 \cdot 10^{ -1 }$ &
\cellcolor{ green!30 }  $ 5.7 \cdot 10^{ -1 }$ & \cellcolor{ green!30 }  $ 7.6
\cdot 10^{ -1 }$ & \cellcolor{ green!30 }  $ 8.8 \cdot 10^{ -1 }$ & \cellcolor{
green!30 }  $ 9.5 \cdot 10^{ -1 }$ & \cellcolor{ green!30 }  $ 9.8 \cdot 10^{ -1
}$ & \cellcolor{ green!30 }  $\approx 1$ & \cellcolor{ green!30 }  $\approx 1$
\\
60 & \cellcolor{ yellow!30 }  $ 6.7 \cdot 10^{ -3 }$ & \cellcolor{ green!30 }  $
4.0 \cdot 10^{ -2 }$ & \cellcolor{ green!30 }  $ 1.2 \cdot 10^{ -1 }$ &
\cellcolor{ green!30 }  $ 2.7 \cdot 10^{ -1 }$ & \cellcolor{ green!30 }  $ 4.4
\cdot 10^{ -1 }$ & \cellcolor{ green!30 }  $ 6.2 \cdot 10^{ -1 }$ & \cellcolor{
green!30 }  $ 7.6 \cdot 10^{ -1 }$ & \cellcolor{ green!30 }  $ 8.7 \cdot 10^{ -1
}$ & \cellcolor{ green!30 }  $\approx 1$ & \cellcolor{ green!30 }  $\approx 1$
\\
80 & \cellcolor{ yellow!30 }  $ 1.3 \cdot 10^{ -3 }$ & \cellcolor{ yellow!30 }
$ 9.8 \cdot 10^{ -3 }$ & \cellcolor{ green!30 }  $ 3.8 \cdot 10^{ -2 }$ &
\cellcolor{ green!30 }  $ 1.0 \cdot 10^{ -1 }$ & \cellcolor{ green!30 }  $ 2.1
\cdot 10^{ -1 }$ & \cellcolor{ green!30 }  $ 3.5 \cdot 10^{ -1 }$ & \cellcolor{
green!30 }  $ 5.0 \cdot 10^{ -1 }$ & \cellcolor{ green!30 }  $ 6.5 \cdot 10^{ -1
}$ & \cellcolor{ green!30 }  $ 9.8 \cdot 10^{ -1 }$ & \cellcolor{ green!30 }
$\approx 1$ \\
100 & \cellcolor{ yellow!30 }  $ 2.4 \cdot 10^{ -4 }$ & \cellcolor{ yellow!30 }
$ 2.2 \cdot 10^{ -3 }$ & \cellcolor{ green!30 }  $ 1.1 \cdot 10^{ -2 }$ &
\cellcolor{ green!30 }  $ 3.4 \cdot 10^{ -2 }$ & \cellcolor{ green!30 }  $ 8.2
\cdot 10^{ -2 }$ & \cellcolor{ green!30 }  $ 1.6 \cdot 10^{ -1 }$ & \cellcolor{
green!30 }  $ 2.7 \cdot 10^{ -1 }$ & \cellcolor{ green!30 }  $ 4.1 \cdot 10^{ -1
}$ & \cellcolor{ green!30 }  $ 9.2 \cdot 10^{ -1 }$ & \cellcolor{ green!30 }
$\approx 1$ \\
120 & \cellcolor{ orange!35 }  $ 4.5 \cdot 10^{ -5 }$ & \cellcolor{ yellow!30 }
$ 5.0 \cdot 10^{ -4 }$ & \cellcolor{ yellow!30 }  $ 2.8 \cdot 10^{ -3 }$ &
\cellcolor{ green!30 }  $ 1.0 \cdot 10^{ -2 }$ & \cellcolor{ green!30 }  $ 2.9
\cdot 10^{ -2 }$ & \cellcolor{ green!30 }  $ 6.7 \cdot 10^{ -2 }$ & \cellcolor{
green!30 }  $ 1.3 \cdot 10^{ -1 }$ & \cellcolor{ green!30 }  $ 2.2 \cdot 10^{ -1
}$ & \cellcolor{ green!30 }  $ 7.9 \cdot 10^{ -1 }$ & \cellcolor{ green!30 }  $
9.9 \cdot 10^{ -1 }$ \\
140 & \cellcolor{ orange!35 }  $ 8.6 \cdot 10^{ -6 }$ & \cellcolor{ yellow!30 }
$ 1.1 \cdot 10^{ -4 }$ & \cellcolor{ yellow!30 }  $ 6.9 \cdot 10^{ -4 }$ &
\cellcolor{ yellow!30 }  $ 3.0 \cdot 10^{ -3 }$ & \cellcolor{ yellow!30 }  $ 9.6
\cdot 10^{ -3 }$ & \cellcolor{ green!30 }  $ 2.5 \cdot 10^{ -2 }$ & \cellcolor{
green!30 }  $ 5.5 \cdot 10^{ -2 }$ & \cellcolor{ green!30 }  $ 1.1 \cdot 10^{ -1
}$ & \cellcolor{ green!30 }  $ 6.1 \cdot 10^{ -1 }$ & \cellcolor{ green!30 }  $
9.7 \cdot 10^{ -1 }$ \\
160 & \cellcolor{ orange!35 }  $ 1.6 \cdot 10^{ -6 }$ & \cellcolor{ orange!35 }
$ 2.3 \cdot 10^{ -5 }$ & \cellcolor{ yellow!30 }  $ 1.7 \cdot 10^{ -4 }$ &
\cellcolor{ yellow!30 }  $ 8.1 \cdot 10^{ -4 }$ & \cellcolor{ yellow!30 }  $ 2.9
\cdot 10^{ -3 }$ & \cellcolor{ yellow!30 }  $ 8.6 \cdot 10^{ -3 }$ &
\cellcolor{ green!30 }  $ 2.1 \cdot 10^{ -2 }$ & \cellcolor{ green!30 }  $ 4.5
\cdot 10^{ -2 }$ & \cellcolor{ green!30 }  $ 4.3 \cdot 10^{ -1 }$ & \cellcolor{
green!30 }  $ 9.2 \cdot 10^{ -1 }$ \\
180 & \cellcolor{ red!40 }  $ 3.1 \cdot 10^{ -7 }$ & \cellcolor{ orange!35 }  $
4.9 \cdot 10^{ -6 }$ & \cellcolor{ orange!35 }  $ 3.9 \cdot 10^{ -5 }$ &
\cellcolor{ yellow!30 }  $ 2.1 \cdot 10^{ -4 }$ & \cellcolor{ yellow!30 }  $ 8.6
\cdot 10^{ -4 }$ & \cellcolor{ yellow!30 }  $ 2.8 \cdot 10^{ -3 }$ &
\cellcolor{ yellow!30 }  $ 7.6 \cdot 10^{ -3 }$ & \cellcolor{ green!30 }  $ 1.8
\cdot 10^{ -2 }$ & \cellcolor{ green!30 }  $ 2.7 \cdot 10^{ -1 }$ & \cellcolor{
green!30 }  $ 8.2 \cdot 10^{ -1 }$ \\ \hline
240 & \cellcolor{ red!40 }  $ 2.1 \cdot 10^{ -9 }$ & \cellcolor{ red!40 }  $ 4.3
\cdot 10^{ -8 }$ & \cellcolor{ red!40 }  $ 4.6 \cdot 10^{ -7 }$ & \cellcolor{
orange!35 }  $ 3.2 \cdot 10^{ -6 }$ & \cellcolor{ orange!35 }  $ 1.7 \cdot 10^{
-5 }$ & \cellcolor{ orange!35 }  $ 7.2 \cdot 10^{ -5 }$ & \cellcolor{ yellow!30
}  $ 2.6 \cdot 10^{ -4 }$ & \cellcolor{ yellow!30 }  $ 7.8 \cdot 10^{ -4 }$ &
\cellcolor{ green!30 }  $ 3.9 \cdot 10^{ -2 }$ & \cellcolor{ green!30 }  $ 3.8
\cdot 10^{ -1 }$ \\ \hline
300 & \cellcolor{ red!40 }  $ 1.4 \cdot 10^{ -11 }$ & \cellcolor{ red!40 }  $
3.6 \cdot 10^{ -10 }$ & \cellcolor{ red!40 }  $ 4.7 \cdot 10^{ -9 }$ &
\cellcolor{ red!40 }  $ 4.1 \cdot 10^{ -8 }$ & \cellcolor{ red!40 }  $ 2.7 \cdot
10^{ -7 }$ & \cellcolor{ orange!35 }  $ 1.4 \cdot 10^{ -6 }$ & \cellcolor{
orange!35 }  $ 6.1 \cdot 10^{ -6 }$ & \cellcolor{ orange!35 }  $ 2.3 \cdot 10^{
-5 }$ & \cellcolor{ yellow!30 }  $ 3.1 \cdot 10^{ -3 }$ & \cellcolor{ green!30 }
$ 9.2 \cdot 10^{ -2 }$ \\ \hline
360 & \cellcolor{ red!40 }  $ 9.4 \cdot 10^{ -14 }$ & \cellcolor{ red!40 }  $
2.9 \cdot 10^{ -12 }$ & \cellcolor{ red!40 }  $ 4.5 \cdot 10^{ -11 }$ &
\cellcolor{ red!40 }  $ 4.7 \cdot 10^{ -10 }$ & \cellcolor{ red!40 }  $ 3.6
\cdot 10^{ -9 }$ & \cellcolor{ red!40 }  $ 2.3 \cdot 10^{ -8 }$ & \cellcolor{
red!40 }  $ 1.2 \cdot 10^{ -7 }$ & \cellcolor{ red!40 }  $ 5.2 \cdot 10^{ -7 }$
& \cellcolor{ yellow!30 }  $ 1.7 \cdot 10^{ -4 }$ & \cellcolor{ green!30 }  $
1.3 \cdot 10^{ -2 }$ \\ \hline
480 & \cellcolor{ red!40 }  $ 4.2 \cdot 10^{ -18 }$ & \cellcolor{ red!40 }  $
1.7 \cdot 10^{ -16 }$ & \cellcolor{ red!40 }  $ 3.6 \cdot 10^{ -15 }$ &
\cellcolor{ red!40 }  $ 4.9 \cdot 10^{ -14 }$ & \cellcolor{ red!40 }  $ 5.0
\cdot 10^{ -13 }$ & \cellcolor{ red!40 }  $ 4.1 \cdot 10^{ -12 }$ & \cellcolor{
red!40 }  $ 2.8 \cdot 10^{ -11 }$ & \cellcolor{ red!40 }  $ 1.7 \cdot 10^{ -10
}$ & \cellcolor{ red!40 }  $ 2.1 \cdot 10^{ -7 }$ & \cellcolor{ orange!35 }  $
8.0 \cdot 10^{ -5 }$ \\ \hline
600 & \cellcolor{ red!40 }  $ 1.9 \cdot 10^{ -22 }$ & \cellcolor{ red!40 }  $
9.8 \cdot 10^{ -21 }$ & \cellcolor{ red!40 }  $ 2.5 \cdot 10^{ -19 }$ &
\cellcolor{ red!40 }  $ 4.3 \cdot 10^{ -18 }$ & \cellcolor{ red!40 }  $ 5.4
\cdot 10^{ -17 }$ & \cellcolor{ red!40 }  $ 5.6 \cdot 10^{ -16 }$ & \cellcolor{
red!40 }  $ 4.7 \cdot 10^{ -15 }$ & \cellcolor{ red!40 }  $ 3.5 \cdot 10^{ -14
}$ & \cellcolor{ red!40 }  $ 1.3 \cdot 10^{ -10 }$ & \cellcolor{ red!40 }  $ 1.8
\cdot 10^{ -7 }$ \\
\end{tabular}
\caption{We display the probability of observing the creation of $n$ or fewer
blocks during the last $t$ minutes, for different values of $n$ and $k$. To
account for natural changes in the hash rate, we assume that the time between
block creations is 12 minutes (instead of 10). The coloring is as described
in~\autoref{sec:alerts}.}
\label{tab:alerts12mins}
\vspace{-4mm}
\end{table*}
\setlength{\tabcolsep}{6pt}

\subsection{Alert Types}\label{sec:alerts}

\newcommand{\colorsquare}[1]{\begin{tikzpicture}[scale=4, every node/.style={transform shape}]\draw[draw=black,fill=#1] (0,0) rectangle (0.05,0.05);\end{tikzpicture}\,}

Using the probabilistic model described above, we introduce \emph{alerts} that
are triggered whenever the probability of a given sequence of block creation
times is below a threshold. Different Bitcoin users require different levels of
reliability, and therefore different thresholds % depending on the value of the
user's typical transactions (see also~\cite{bitcoinj}). A user who cares more
about fast confirmation times than security, e.g., an automated online store for
video game downloads, might accept a transaction if it appears on \emph{any}
main chain block, with no confirmations needed. However, users with high
security expectations, e.g., financial service providers, are more likely to use
the \emph{six-confirmations rule}, which means that they will only accept a
transaction if it appears in a block that has six consecutive main chain blocks
built on top of it.

In~\autoref{tab:alerts12mins}, we have displayed for a range of combinations of
$t$ and $k$, the probability of observing $k$ blocks during a time period of $t$
minutes. These probabilities were computed using the statistical package
R.\footnote{\url{https://www.r-project.org/}} The probabilities have been
color-coded in the following way: a cell in~\autoref{tab:alerts12mins} is
colored \emph{yellow}
\colorsquare{yellow!30}, \emph{orange} \colorsquare{orange!35}, or \emph{red}
\colorsquare{red!40} if the corresponding probability is lower than $10^{-2}$,
$10^{-4}$, or $10^{-6}$ respectively --- the cell is colored \emph{green}
\colorsquare{green!30} if it neither. To compare this to certainty thresholds
that are common in the scientific literature: for hypothesis testing in social
science and clinical trials and threshold of $5\%$, i.e., $5\cdot 10^{-2}$ is
used, whereas the `\emph{five sigma}' rule used to mark a discovery in physics
correspond to $2.87 \cdot10^{-7}$. Although the choice of thresholds for the
alert types is necessarily subjective, the yellow and red flag thresholds were
chosen to be roughly between these values. Yellow flags might occur relatively
even under normal circumstances: since 144 blocks are mined per day, an average
1.44 blocks per day will be have a creation time that is in the top $1\%$. If
the number of false alarms is too high, then this by itself may be inconvenient
to the user --- in any case, the exact alert level at which a user is
comfortable accepting a transaction depends on her preferences.

In practice, we recommend the following scheme. We aim to detect two types of
eclipse attacks: 1) an attacker without any significant mining power on her
chain performing a DoS attack, and 2) an attacker with significant mining power
who aims to double-spend. Regarding 1), the lightweight client tracks how much
time has passed since the creation time (indicated by the timestamp) of the last
block. If this is longer than approximately 55, 110, or 165 minutes, then this
throws a yellow, orange, or red flag, respectively (these values are derived
directly from the exponential distribution). For attack 2), we keep track of the
creation times of the last $k+1$ blocks, where $k$ is the number of
confirmations. If $k=6$, then this is longer than approximately 190, 275, or 350
minutes, then this throws a yellow, orange, or red flag, respectively (these
values are derived from the Erlang distribution). To summarize, although one
would normally except 7 blocks to be mined after 70 minutes, one should get
suspicious if it took three hours, and if it took six hours one can say with
near-certainty that an attack is taking place.

\subsubsection*{Timestamp Reliability}
We have so far assumed that the timestamps are reliable: our analysis shows that
under normal circumstances this is a reasonable assumption, even though small
deviations are tolerated by miners and lightweight
clients~\cite{szalachowski2018towards}. In $0.5\%$ of all blocks observed in the
study period, the timestamp was \emph{lower} than the timestamp of the previous
block, although these differences were often less than 30 seconds. These
observations were discarded to create~\autoref{fig:exponential_qq}. Deviations
may occur because even honest miners have considerable freedom when choosing the
block timestamps. For example, the miner could choose to fix the timestamp when
she starts to mine a new block, or update the timestamp continuously while
mining. If different miners follow different rules, then this may lead to
deviations from the exponential distribution. Furthermore, network latency adds
a period of time with an unclear probability distribution between the creation
of a block and the start of the mining process for the next block. Still, our
analysis of real block timestamps suggests that these effects are negligible
compared to the roughly 10 minutes on average between block arrivals, and that
they do not have a major effect on the shape of the tail (i.e., the likelihood
that very large values occur). 

During an attack, the attacker is also free to set the timestamps at will. This
has no effect if the attacker is performing a DoS attack, because in that case
the attacker is not mining blocks. However, if she is trying to double-spend,
then she can choose the timestamps such that the blocks appear to have been
mined earlier. To ameliorate this, the client can also consider the times at
which the blocks are first observed \textit{by the client} instead of the
timestamps in the block headers. Let $\Delta$ denote the difference between the
arrival time of the last block and the client's current system clock. The client
can then add $\Delta$ to the total time for the seven blocks -- after all, this
is a lower bound on the last block's contribution  to the total. Again, there
are some concerns regarding the effect of the network latency on the probability
distribution of the times between block arrivals. We leave the evaluation of
whether the exponential distribution is still a good fit in this setting as
future work.

\subsection{Attack Analysis}

In~\autoref{tab:attack_probs}, we have displayed for several values of
$\alpha$ the probability that an $\alpha$-strong is able to create 7 blocks
without triggering an alert, for each of the four alert types. We assume that
the attacker controls at most 50\% of the mining power, as otherwise it would be
easier to perform a $51\%$ attack on the whole network without eclipsing
specific clients. We see that even for a $20\%$-strong attacker, the probability
of creating 7 blocks without of triggering a yellow alert is less than $1.7\%$.
For a $5\%$-strong attacker, the probability of not triggering a red alert is
already less than $0.01\%$. This means that an attacker will need considerable
mining power to perform double-spend attacks without the risk of causing alarm,
even when simultaneously performing an eclipse attack.

\begin{table}[!h]
\centering
\small
\begin{tabular}{cccc} 
$\alpha$ & \emph{yellow} \colorsquare{yellow!30} & \emph{orange}
\colorsquare{orange!35} & \emph{red} \colorsquare{red!40} \\ \midrule 0.05 & $
2.05 \cdot 10^{ -6 }$ & $ 2.72 \cdot 10^{ -5 }$ & $ 1.40 \cdot 10^{ -4 }$ \\
0.08 & $ 5.78 \cdot 10^{ -5 }$ & $ 6.40 \cdot 10^{ -4 }$ & $ 2.82 \cdot 10^{ -3
}$ \\
0.125 & $ 1.10 \cdot 10^{ -3 }$ & $ 9.34 \cdot 10^{ -3 }$ & $ 3.28 \cdot 10^{ -2
}$ \\
0.2 & $ 1.68 \cdot 10^{ -2 }$ & $ 9.44 \cdot 10^{ -2 }$ & $ 2.33 \cdot 10^{ -1
}$ \\
0.3 & $ 1.13 \cdot 10^{ -1 }$ & $ 3.85 \cdot 10^{ -1 }$ & $ 6.46 \cdot 10^{ -1
}$ \\
0.5 & $ 5.47 \cdot 10^{ -1 }$ & $ 8.85 \cdot 10^{ -1 }$ & $ 9.77 \cdot 10^{ -1
}$ \\
\end{tabular}
\caption{We display the probability that an $\alpha$-strong attacker is able to
create 7 blocks within a period that is short enough to not trigger an alert,
for each of the 3 alert types.}
\label{tab:attack_probs}
\vspace{-4mm}
\end{table}

\section{Gossip-Based Protocol}
\label{sec:gossipoverview}
In this section,~\autoref{subsec:gossipoverview} discuss the overview of the
gossip-based protocol,~\autoref{sec:ProtocolForBitcoin} give protocol
description of passive-based gossiping approach and~\autoref{sec:analysis}
provides a detailed analysis of its effectiveness. Last,~\autoref{sec:active}
shortly discuss an active-based gossiping approach which helps to further
improve detection time. The notation used to describe our gossip-based protocol
is presented in~\autoref{table:notations}. 
\begin{table}[h]
\small
\begin{tabular}{llr}
\hline
Notation & Description \\ \hline
$S$  & a server participating in the protocol\\
$LC$ & a (light) client  participating in the protocol\\
{$V_{LC}$} & a blockchain view of {$LC$}  \\ 
{$V_{S}$} & a blockchain view of {$S$}  \\ 
$HDR_{S->LC}$ & a set of block headers send from $S$ to $LC$ \\ 
$HDR_{LC->S}$ & a set of block headers send from $LC$ to $S$ \\ 
\hline
\end{tabular}
\caption{Summary of the used notations.}
\label{table:notations}
\end{table}

\subsection{Protocol Overview}
\label{subsec:gossipoverview}
Bitcoin is a trustless, decentralized network and
our approach for detecting eclipse attacks follows these core principles.  The
proposed protocol does not implicitly need to trust any one server in the gossip
network. Rather, it assumes that if a client can connect to at least one server
with a legitimate view of the blockchain, then a potential eclipse attack can be
detected.  In our approach client-server connections are driven by the user's
natural Internet traffic. For every such connection, a client piggybacks its
blockchain view to the server, which in turn returns its strongest view.
Afterwards, the communicating parties can update their views according to the
strongest chain rule. In such an approach, servers are passive and not connected
to the Bitcoin network, but they maintain their strongest blockchain views based
on the input from clients.

\begin{figure}
	\centering
	\def\svgwidth{\columnwidth}
	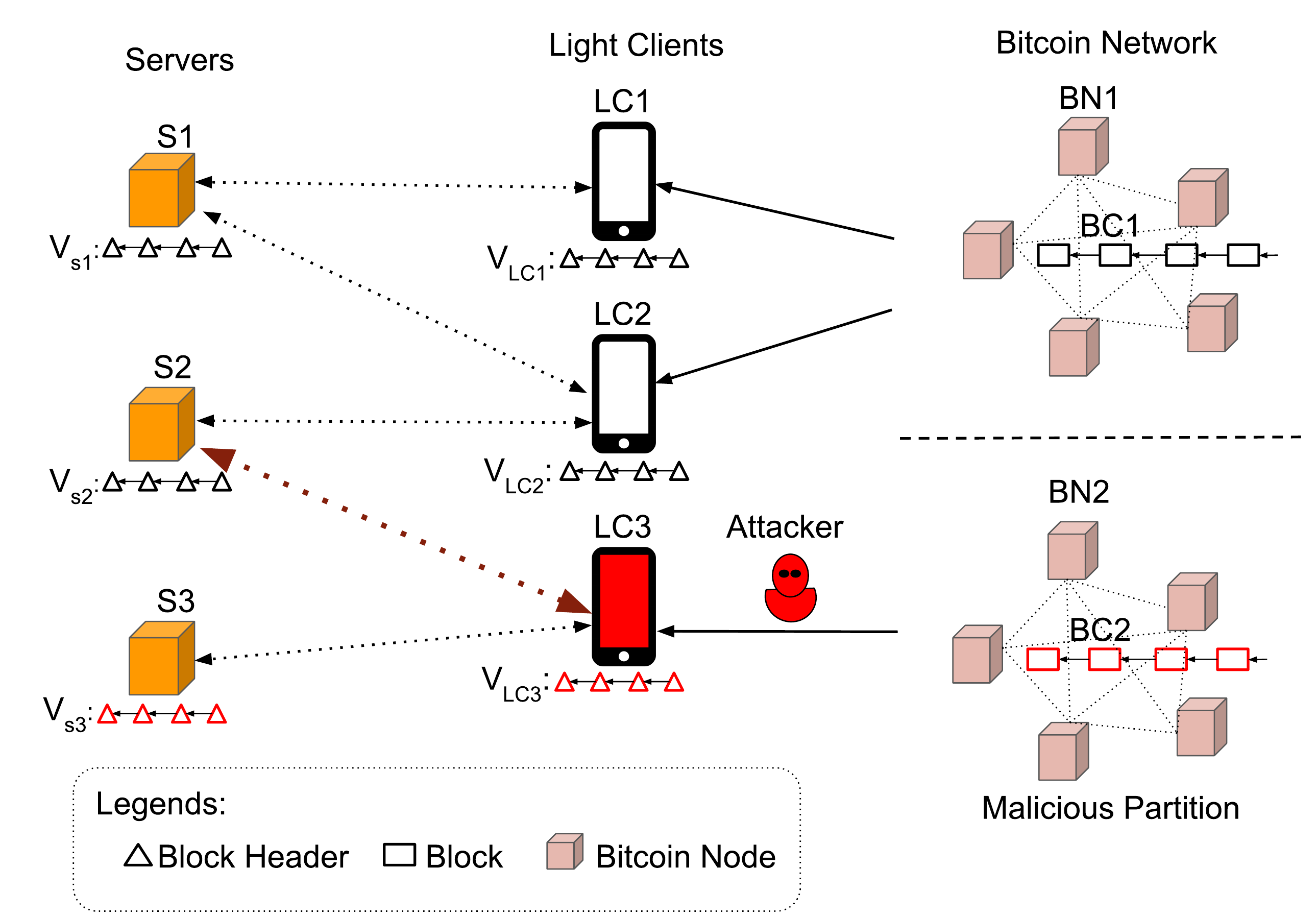
	\caption{Proposed eclipse attack detection model comprising of
		servers, light clients, the Bitcoin network, and the attacker. Blocks/headers that are colored red were created by the attacker.}      
	\label{fig:architecture}
	\vspace{-2mm}
\end{figure}

To illustrate the process better, we depict the main elements and intuitions
behind our framework in \autoref{fig:architecture}. Let us consider the
selected case where there are three servers and three clients supporting our
scheme. Clients $LC_1$ and $LC_2$ are connected to the genuine Bitcoin network
$BN_1$ with the $BC_1$ blockchain view (i.e., headers), while the client $LC_3$
is under an eclipse attack and connected to the malicious $BN_2$ with the view
$BC_2$.  The agents communicate in the following sequence:

\begin{enumerate}
    \item The clients {$LC_1$} and {$LC_2$} holding genuine blockchain views
        contact the server $S_1$, which after their connections validates and
        accepts  the received (authentic) view. 
    \item {$LC_3$} is eclipsed and connected to a malicious Bitcoin network
        partition {$BN_2$}, and while connecting to the server $S_3$ (without any
        current view), the malicious view is accepted by the server.
    \item Afterwards, the attacked client {$LC_3$} connects to the server $S_2$.
        The server and client exchange their blockchain views.
    \item The server {$S_2$} stores the received headers from the client
        $LC_3$. It validates the view received from $LC_3$ and
        picks the chain that corresponds to the strongest view (which is
        expected to be part of the canonical blockchain).
        The server sends its (strongest) view to the client.
    \item {$LC_3$} is now able to detect that it is under an eclipse attack just 
        by concluding that there exists a stronger blockchain than the one obtained 
        from the malicious partition.
\end{enumerate}

We emphasize that in \textit{passive mode} gossiping, the detection process is
driven by natural Internet traffic as participating servers are assumed to
already exist (e.g., running web services), and the protocol messages are
piggybacked on the standard client-server connections. In the case of
\textit{active mode} gossiping, we assume that clients make explicit connections
with known servers to compare their view of the blockchain with that of the
servers before when they check their wallet.

\subsection{The Passive Gossip-Based Protocol}
\label{sec:ProtocolForBitcoin}
In \autoref{sec:baseline}, we saw that based on timestamps alone, a user can at
best get suspicious after an hour and be almost certain of an attack after
roughly three hours. If the attacker controls some mining power on her branch,
then attack detection is even slower. In this section, we propose a gossip-based
protocol to detect eclipse attacks more rapidly. It is described as a series of
interactions between the server, lightweight client, and the Bitcoin network.
The client$-$server communication is part of the natural client's traffic
whereas the client$-$Bitcoin network communication is on demand, like today. The
main goal of the client$-$server communication is to provide a client, via a
server, with the strongest view of the chain the server has seen. To achieve
this, a lightweight client forwards to a contacted server headers obtained from
the Bitcoin network. When a  server communicates with multiple clients, the
server stores the strongest view of the chain it has received and serves this
view to connecting clients. As a result, a client viewing a malicious partition
of the Bitcoin network is able to fetch the strongest chain seen by the server
and compare with its own. When the server has a stronger view of the chain, the
client can conclude that it is under an eclipse attack. In addition to detecting
eclipse attacks, we propose a Bitcoin header size optimization to reduce
communication overheads. We choose the headers of HTTP(S) request/response
messages as a communication medium of our gossip layer due to the ubiquitous
nature of HTTP(S). Furthermore, if HTTPS is used then the message headers are
encrypted, which means that a man-in-the-middle attacker cannot distinguish
these messages from normal traffic. We discuss the implications of our choice
in~\autoref{sec:gossip}. We focus on the passive version of the protocol in this
section, and shortly discuss the active version in~\autoref{sec:active}.  

 \subsubsection*{Storage Requirement}
 The storage on the client and the server is a fixed-length sequence of Bitcoin
 headers, referred to as a queue or window. New headers enter at the
 \textit{tail} of the queue and old headers exit at the \textit{head}. The
 client and server windows are of equal size. As we explain later, the maximum
 storage requirement on a client/server is 2016 times 80 bytes, i.e., around 160
 kB and we can also compress it as mention
 in~\autoref{sssec:HeaderSizeReduction} to reduce storage requirement by half.
 Here, 2016 corresponds to the average number of Bitcoin blocks generated in two
 weeks and 80 bytes is the Bitcoin header size.
  
 \subsubsection*{Network Interactions}
 The main agents of our protocol are the server(s), client(s), and the Bitcoin
network. A client may have newly joined, or been offline for a period of time.
The first step for the client is to get its headers up-to-date from the Bitcoin
network. It may, however, get updated with  malicious headers when it is under
an eclipse attack. The second step is the exchange of headers between the client
and a protocol-running server via HTTP(S) messages. However, a server that
starts to deploy our protocol may have an empty window and no headers to send.
If this is the case, it accepts the headers provided by the client and returns
an empty (NULL) window to the client. The message exchange that follows is
discussed in~\autoref{sssec:MessageExchange}. The third step is to match the
views on the client and server. This is discussed in~\autoref{sssec:ViewMatch}.
The final step is to find the strongest chain, as we discuss
in~\autoref{sssec:FindingStrongectChain}. ~\autoref{sssec:HeaderSizeReduction}
presents an optimization technique for the block headers to save both storage
and bandwidth. ~\autoref{sec:gossip} discusses the choice of layer used for
message exchange.

\subsubsection{Message Exchange Between Client and Server}
\label{sssec:MessageExchange}

\begin{figure}[b!]
  \centering
  \def\svgwidth{\columnwidth}
  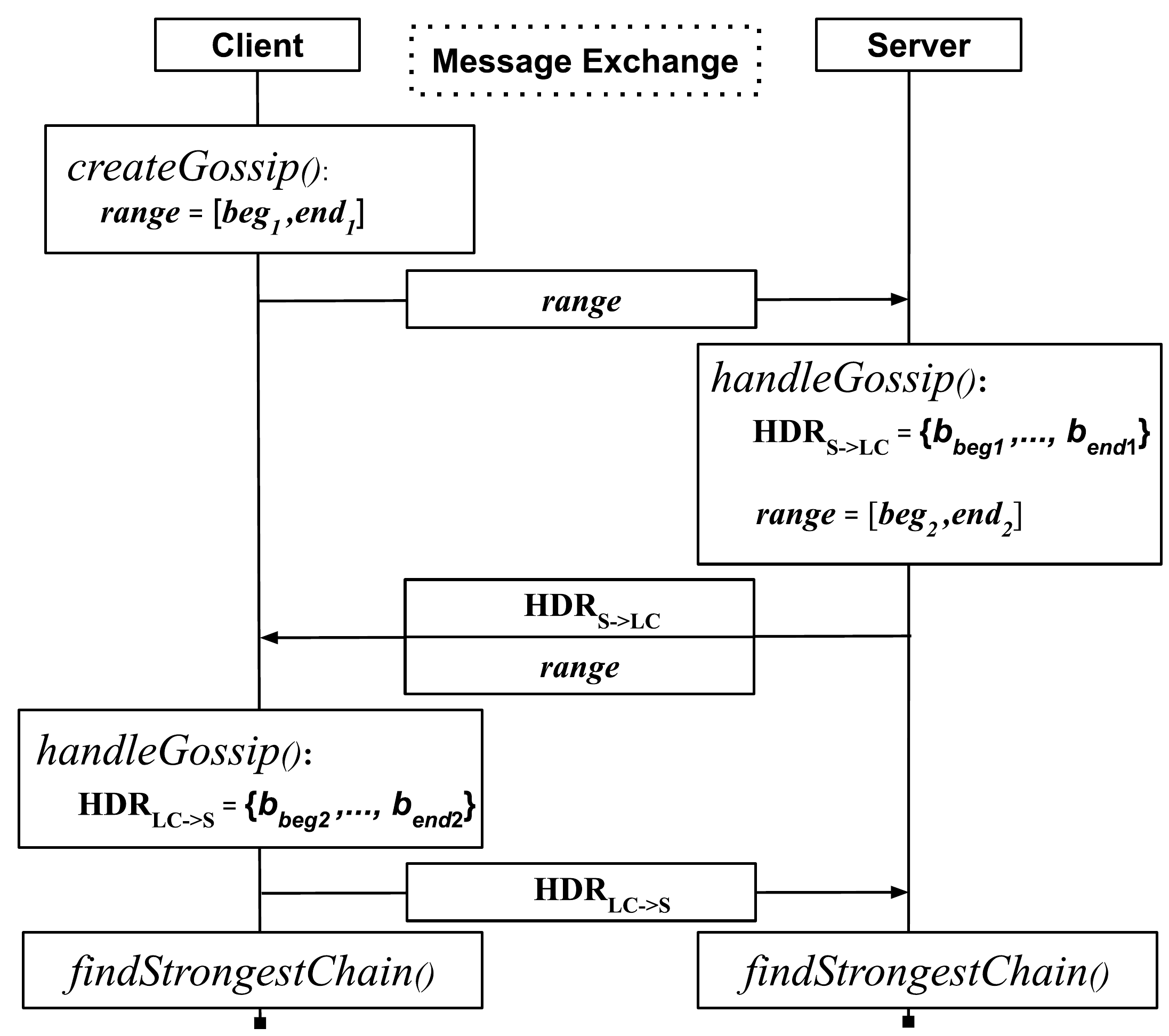   
  \caption{Header exchange between client and server.}       
  \label{fig:handshake}
\end{figure}
 
In the following, we assume that a client has a copy of the headers from the
Bitcoin network, but is unable to ascertain whether they belong to an inferior
chain created by an attacker. Hence, the client wishes to connect to a server or
group of servers and confirm the veracity of the headers it received via its
natural HTTP(s) connections. The message exchange between the client and a
server begins with a service request from the client.

\autoref{fig:handshake} outlines the message exchange between the client and the
server. We note that the header index of the genesis block is 0 and we define
\textit{last} as the index of the latest header. This would ensure that both
chains are compared over the same index range avoiding offset calculation
mistakes (please note that Bitcoin headers do not contain any sequence numbers).
Let $range=[beg,end]$ be the sequence of blocks between a given beginning and
ending index, respectively. The messages exchanged are $\romannumeral 1$) a
request for headers in a range or $\romannumeral 2$) a set of headers. 
 
The gossip message created by the client is a \textit{range} of headers and is communicated to the server which
handles the gossip message received. 
The set of requested headers ($HDR_{S->LC}$) and a new \textit{range} are 
sent to the client, who responds to the server with the requested headers ($HDR_{LC->S}$).
The client and the server have now completed their exchange of headers.

\begin{algorithm}
    \SetKwFunction{Size}{Size}
    \SetKwFunction{Append}{Append}
    \SetKwInOut{Input}{input}
    \SetKwInOut{Output}{output}
	\SetKwProg{Fn}{function}{}{}
    \Input{{${V}_{LC}$},{$V_{S}$}; {${V}_{LC}$} $\neq$ {$V_{S}$} }
    \Output{\textit{strongestChain}}
    \BlankLine
  \Fn{findStrongestChain ({${V}_{LC}$},{$V_{S}$})}{
    $strongestChain \leftarrow$ \{\} \;
    $sWeight \leftarrow 0$,	$cWeight \leftarrow 0$ \;
	
    \BlankLine
   \ForEach{{${b}$} {$\in$} {${V}_{LC}$}}{
      $targetHash\leftarrow find\_target({b})$ \;
    $cWeight\leftarrow cWeight + \textit{targetHash} $ \;
    }

    \BlankLine
   \ForEach{{b} {$\in$} {${V}_{S}$}}{
      $targetHash\leftarrow find\_target(\textit{b})$ \;
    $sWeight\leftarrow sWeight + \textit{targetHash} $ \;
    }
	
	\BlankLine	
	\If(){$cWeight>sWeight$}{
          $\textit{strongestChain} \leftarrow$ {$V_{S}$}\;
        } 
	\ElseIf(){$cWeight<sWeight$}{
	  $\textit{strongestChain} \leftarrow$ {${V}_{LC}$}\;
			}
    
    \BlankLine	
    \Return ${strongestChain}$\;
 }
    \caption{Find $strongest$ chain given 2 views}
    \label{alg:alg_strongestchain}
  \end{algorithm}

\subsubsection{View Matching}
\label{sssec:ViewMatch}

Once the messages have been exchanged, the next step is to compare them.
However, it needs to be ensured that comparisons are over the same range of
headers. To achieve this goal, we formalize the view on the server and the
client. The server view is its window of the latest headers: {$V_{S}$} =
\{{$b_{n+1}$},...,{$b_{n+k}$}\}. The client view is 
\mbox{{${V}_{LC}$} = \{{$\hat{b}_{n+1}$},...,{$\hat{b}_{n+k}$}\}. }
Note that the server and client exclude headers received from their most recent
message exchange. Hence, we have, ${V}_{S} \leftarrow$ {${V}_{S}\setminus b;
{\{b:b\in HDR_{LC->S}, b\notin {V}_{S}\}} $} and ${V}_{LC} \leftarrow$
{${V}_{LC}\setminus b; {\{b:b\in HDR_{S->LC}, b\notin {V}_{LC}\}} $}. The
excluded headers from their current view are added after the strongest chain is
computed (see \autoref{sssec:FindingStrongectChain}). This ensures that
comparisons are not made on the same copy of headers exchanged from one side to
the other.

When no forks are detected, the server and the client are viewing the same set
of headers and {$b$}={$\hat{b}$}. When a fork is detected at {$b^{*}$}$\in$
{$V_{S}$}, {${V}_{LC}$}; the subsequent {$b$}{$\neq$}{$\hat{b}$}. Here, the
server and client are viewing the same set of header up to {$b^{*}$}, and the
subsequent views diverge. The fork is resolved by following the Bitcoin's
\textit{strongest} chain applied to views. All message exchanges in a window of
size \textit{k}, are bounded to 2$*$\textit{k} header transfers, for one round
of client$-$server communication. The ability of the client to detect and
resolve a fork is limited to forks occurring within the window. For instance, a
window size of \mbox{$k=2016$} corresponds to roughly 2 weeks (for an average
inter-block time of 10 minutes) and a client who is offline for this period is
still able to detect a fork. The size of the window is a system parameter and
may be increased to support a longer detection period.

\subsubsection{Finding the Strongest Chain}
\label{sssec:FindingStrongectChain}
Once the views are matched, the client and server are ready to find the
\textit{strongest} chain. By the nature of the message exchange, clients and
servers may see different views of the chain over a period of time (within the
same range). When presented with two views, they pick the stronger chain as
specified by the Bitcoin protocol (note that any stronger chain presented can
convince an attacked client that she is under the attack). This process is
repeated for every round of a message exchange and the full algorithm is
described in ~\autoref{alg:alg_strongestchain}, taking the views {$V_{S}$} and
{$V_{LC}$} as input and outputting the \textit{strongest} chain. Lines 4-7 and
8-11 calculate the cumulative sum of the proof-of-work of the headers in
{$V_{LC}$} and {$V_{S}$} respectively (which is derived from
\textit{nBits}~\cite{nbitstarget} for each header). Lines 12-17 compare the
cumulative sum to decide on the \textit{strongest} chain. Recall that for a
smaller target hash, a higher hash rate is required to solve a proof-of-work
puzzle. As a result, the smaller cumulative sum determines which view is
stronger. Once the strongest chain is determined at the server, all ${\{b:b \in
HDR_{LC->S}, b \notin V_{S}\}}$ that can form a valid chain are added to
$V_{S}$. Similarly, at the client, all valid ${\{b:b \in HDR_{S->LC}, b \notin
V_{LC}\}}$ are added to $V_{LC}$. Note that the window is implemented as a
fixed-length FIFO queue. 
The last step of extending the view is to prepare the protocol for the next
round of communication. Headers are not checked if they are part of the strongest
chain seen yet.

\subsubsection{Header Size Reduction}
\label{sssec:HeaderSizeReduction}

As our protocol requires that blockchain fragments (i.e., consecutive block
headers) are sent between clients and servers, we propose a way of minimizing this
overhead. The main intuition behind our modification is that the
\textit{prevHash} field of the Bitcoin header can be computed from the previous
header. Therefore, for a subchain consisting of consecutive block headers, only
the first header has to include its \textit{prevHash}, whereas every subsequent
header can compute it recursively. Another observation is that some header
fields have either constant or infrequently changing values.

\begin{figure}[h!]
  \centering
  \def\svgwidth{\linewidth}
  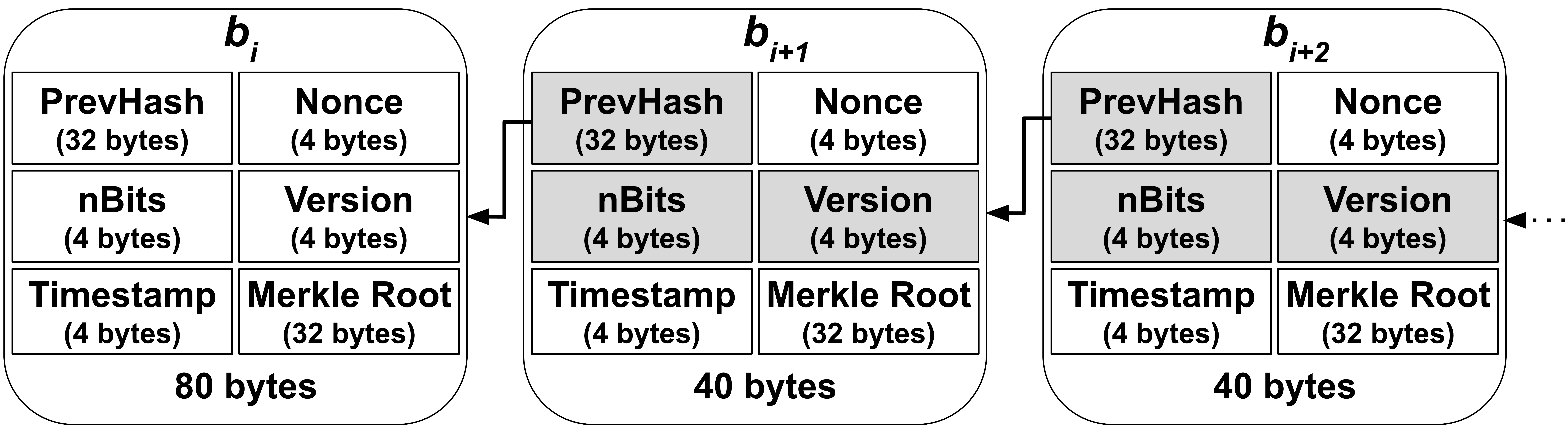   
   \caption{Header size reduced by removing \textit{version}, \textit{nBits} and \textit{prevHash} fields.}      
  \label{fig:blockcompression}
\end{figure}
In~\autoref{fig:blockcompression}, we present a header list where the full
header is only stored in $b_i$, whereas $b_{i+1}$ and $b_{i+2}$ have their
highlighted fields removed. While each Bitcoin header is 80 bytes in size, it is
unnecessary to store all fields as \textit{nBits} is fixed for 2015 consecutive
blocks, \textit{version} changes infrequently, and \textit{prevHash} can be
recalculated as described above. Hence, in our protocol we remove \textit{nBits}
(we send it only for difficulty changes -- every 2016 blocks), \textit{version}
(we send it only for version changes), and \textit{prevHash} fields from most
headers that are part of the message exchange. These changes help us to reduce
the size of a typical header from 80 to only 40 bytes.

\subsubsection{Gossip Layer}
\label{sec:gossip}
As mentioned before, we have implemented our protocol by appending the block
headers to the header of an HTTP request-response. This facilitates the easier
deployment of our protocol as no higher-level protocol modifications are
required. Also appending additional data into the HTTP header does not break
applications not supporting our protocol since server drops any unnecessary data
in the header without affecting the protocol. Though HTTP protocol specification
does not specify any limit on the amount of data that can be sent in a HTTP
headers, web servers usually restrict the size to 4k-64k. However, this limit is
configurable and can be set according to the requirement of our protocol.
Although selecting HTTP as gossiping medium make our protocol application
specific means exclude non-HTTP(S) traffic that can also be leveraged to gossip,
we have found that in practice, HTTP(S) dominates Internet traffic. In fact,
non-HTTP(S) traffic only accounts for 12\% of Internet traffic in our dataset.
This is discussed in more detail in~\autoref{sec:analysis}. We note that
different choices for the gossip layer could be made, e.g., the HTTP(S) body
could be modified instead, or TCP headers could be used to capture a broader
category of traffic. However, TCP header modification could have unforeseen
effects on middleware that would undo any advantages. For completeness, we
present implementations of gossip medium with TCP
in~\autoref{appendix:gossiptcp} to make our scheme application-agnostic and HTTP
body in~\autoref{appendix:http}.

\subsection{Analysis of the Passive Gossip Protocol}
\label{sec:analysis}
\newcommand{\serverfrac}{p_s}
\newcommand{\clientfrac}{p_c}
\newcommand{\E}{\mathbb{E}}

In this section, we analyze real Internet traffic trace to demonstrate that our
approach of \textit{passive mode} gossiping is practical and efficient compared
to the timestamp-based protocol of ~\autoref{sec:baseline}. In fact, a user may
be able to detect with certainty that she is under attack in roughly one hour
compared to 3 hours previously. We first discuss the data in~\autoref{sec:data}.
We discuss our exact evaluation metrics --- coverage of client IPs, speed of
attack detection, and server freshness --- in~\autoref{sec:methodology}. We
evaluate these metrics using the data in~\autoref{sec:results}.

\subsubsection{Description of Data} \label{sec:data}
To understand the Internet access pattern of users we analyze the traffic log
from a university firewall\footnote{We do not disclose the university name to
comply with the double-blind review process.}. The users authenticate to the
firewall before they access the Internet. This allows us to obtain access
details of their devices to public IP addresses from the firewall traffic log.
As the user devices are assigned IPs through DHCP, we identify users by their
user name rather than their DHCP-assigned IP address\footnote{In accessible
traffic datasets, users are not identifiable. Hence, it is necessary to assume
the user:IP mapping. Our dataset allows to associate users with connections, to
improve the accuracy of our study and to make it realistic.}. In some cases,
users access public servers within the university network through a private IP,
hence we also include those IP addresses to analyse our protocol. In order to
include only traffic in which we can gossip (via HTTP(S) message bodies), we
make use of the \textit{port} field in the traffic log. We also consider
connections which were blocked by the firewall, since we assume that these
connections would have succeeded outside the university.

\begin{figure}[t!]
  \centering
  \def\svgwidth{\columnwidth}
  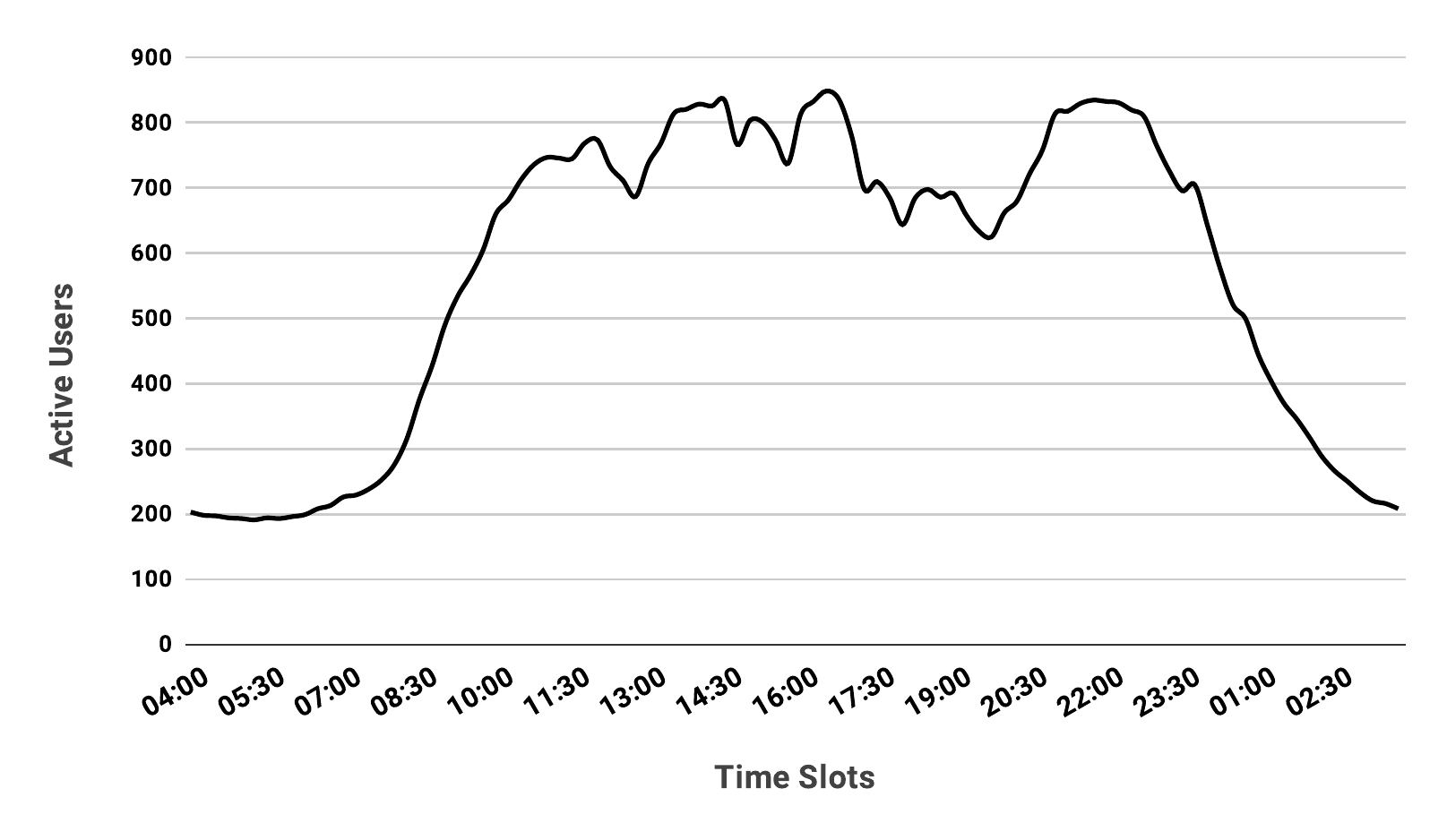   
  \caption{The traffic logs of 24hrs are divided into 15 minutes time-slots and average of total traffic 
  in each slot across the 72hrs traffic log is represented in the graph.}     
  \label{fig:access}
  \vspace{-1mm}
\end{figure}

In total, we collected 72 hours of continuous firewall traffic log. This
included 229,374 unique client IPs (assigned using DHCP) and 269,069 accessed
server IPs. However, these client IPs were mapped to 2511 users and we selected
43277 server IPs that were active in at least 4 out of 12 epochs (the total
72-hour time period was divided into 12 epochs). In order to understand the user
activity, every 24 hours (representing a day) was divided into 15-minute slots.
The average number of connections for each slot across the 72 hrs of traffic log
is represented in~\autoref{fig:access}. We can see that user activity is high
during certain time intervals (active periods) and less so during other time
intervals (inactive periods) --- we assume that many users are sleeping during
the inactive period. Later, we explain how these inactive periods influence our
results. 

\subsubsection*{Ethical Consideration} The traffic log was provided to carry out
research without breaching any user privacy policy. We obtained the following
details: timestamp, server IP, client IP, action and a unique string
representing the user's ID. No other details regarding the users were used for
analysis. We do not publish the dataset with this paper, and as such its use
constitutes \textit{minimal risk} to the users.

\subsubsection{Methodology} \label{sec:methodology}
Given the full dataset, let $\mathcal{U}$ be the full set of users, and
$\mathcal{S}$ the full set of server IPs observed. Furthermore, let $t_0$ and
$t_{\max}$ be the times of the first and last connections respectively, and
$\mathcal{T} = [t_0, t_{\max}]$. We can then extract from the dataset the full
set $\mathcal{C}$ of connections, where each connection $c \in \mathcal{C}$ is a
tuple $(u_c, s_c, t_c)$, such that
\begin{itemize}
\item $u_c \in \mathcal{U}$ is the ID of the user who initiates the connection,
\item $s_c \in \mathcal{S}$ is the IP address of the server, and
\item $t_c \in \mathcal{T}$ is the initiation time of the connection. 
\end{itemize}
We then use $\mathcal{C}$ to define the following metrics for the performance of
the gossip-based approach:

{\bf Coverage:} given a set $S \subset \mathcal{S}$ of servers, we define its
coverage as the fraction of unique users that at some point connect to a server
in $S$. This can be formally expressed as
$$
\text{Coverage}(S) = \frac{|\left\{u \in \mathcal{U}: \exists c \in \mathcal{C} \;\text{ such that }\; s_u  \in S\right\}|}{|\mathcal{U}|},
$$
where $|A|$ denotes the number of elements in set $A$.

{\bf Attack Detection Speed:} given a user $u \in \mathcal{U}$ and a set $S
\subset \mathcal{S}$ of servers, we define the \emph{attack detection speed} as
the time until the next connection from $u$ to a server in $S$. To make this
formal, we first define $C_{u,S}$ as the set of all connections to $S$ initiated
by $u$, i.e., 
$$
C_{u,S} = \{c \in \mathcal{C} : u_c = u \text{ and } s_c \in S\}.
$$ 
For any $t \in \mathcal{T}$, let $T_{u,S}(t)$ be the set of all time points
\emph{after} $t$ at which $u$ connects to a server in $S$. That is,
$$
T_{u,S}(t) = \{ t' \in \mathcal{T} : \exists c \in C_{u,S} \;\text{ such that }\; t_u = t' \text{ and } t_u > t\}.
$$ 
We then define $\delta_{u,S}(t)$ as the time from $t \in \mathcal{T}$ until
$u$'s next connection to a server in $S$ (or until $t_{\max}$). That is,
$$
\delta_{u,S}(t) = \left\{ \begin{array}{cl} \min \left(T_{u,S}(t)\right) - t & \text{if } T_{u,S}(t) \neq \emptyset, \\ t_{\max} - t & \text{otherwise.} \end{array} \right.
$$ 
(Note that $T_{u,S}(t) \neq \emptyset$ only for $t$ after the time of the last
connection, at which point we consider a connection to occur exactly at the end
of the observation period.) We will assume that attacks occur at a time that is
drawn uniformly from $\mathcal{T}$.  The Average Attack Detection Time (AADT)
for $u$ to $S$ is then given as follows:
\begin{equation}
\text{AADT}(u,S) = \frac{1}{t_{\max} - t_{0}} \int_{\mathcal{T}} \delta_{u,S}(t) dt \label{eq:aadt}
\end{equation}
The reasoning behind \eqref{eq:aadt} is as follows. Let the time at which the
attack occurs be given by the random variable $T^*$, and its probability density
function by $f_{T^*}(t) : \mathcal{T} \rightarrow [0,\infty)$. It then
holds\footnote{See, e.g.,\url{https://en.wikipedia.org/wiki/law_of_the_unconscious_statistician}} that
\begin{equation}
\E(\delta_{u,S}(T^*)) = \int_{\mathcal{T}} \delta_{u,S}(t) f_{T^*}(t) dt, 
\label{eq:aadtderv}
\end{equation}
where $\E$ denotes the expected value of a random variable. Substituting
$\frac{1}{t_{\max} - t_{0}}$ for $f_{T^*}(t)$ (which follows from the uniform
distribution of the attack times) into \eqref{eq:aadtderv} then leads to
\eqref{eq:aadt}.

\begin{figure}[!t]
\includegraphics[width=0.48\textwidth]{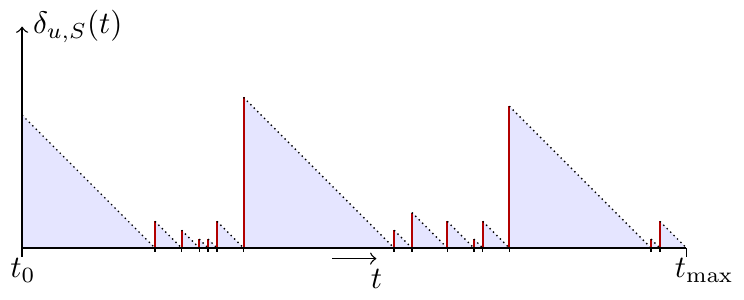}
\caption{Graphical illustration of the average attack detection times given a
user $u$ and a set of servers $S$. Connections from $u$ to a server in $S$ occur
at the time points indicated by the vertical red bars. The dotted lines
represent $\delta_{u,S}(t)$ --- the time of the next connection after $t$.}
\label{fig:averages}
\vspace{-1mm}
\end{figure}

A graphical representation of the AADT is given in~\autoref{fig:averages}. In
this example, the connections occur at the locations of the vertical red bars,
and the height of the red bar indicates the amount of time until the next
connection. The function $\delta_{u,S}$ is represented by the dotted lines, and
the area of the blue triangles represents the integral in~\autoref{eq:aadt}.

{\bf Server Freshness:} given a server $s \in \mathcal{S}$ and a set $U \subset
\mathcal{U}$ of users, we define the \emph{freshness} as the time since the
previous connection from a user in $U$ to $s$. It can defined in a similar way
as the attack detection time. That is, let $T'_{s,U}(t)$ be defined as the set
of all time points \emph{before} $t$ at which a user in $U$ connects to $s$.
Then $\eta_{s,U}(t)$ is the amount of time since the last connection by a user
in $U$ to $s$, defined as follows:
$$
\eta_{s,U}(t) = \left\{ \begin{array}{cl} t - \max \left(T'_{s,U}(t)\right) & \text{if } T'_{s,U}(t) \neq \emptyset, \\ t - t_{\min} & \text{otherwise.} \end{array} \right.
$$ 
The average server freshness can then be defined as follows:
\begin{equation}
\text{Freshness}(s,U) = \frac{1}{t_{\max} - t_{0}} \int_{\mathcal{T}} \eta_{s,U}(t) dt \label{eq:freshness}
\end{equation}
A graphical representation of $\eta_{s,U}(t)$ would look similar
to~\autoref{fig:averages}, but with the triangles flipped horizontally.

\subsubsection*{Active/Inactive Periods} As can be seen from
the~\autoref{fig:averages}, long periods of inactivity --- represented by the
three large triangles --- have a considerable impact on the AADT. One typical
period of inactivity is the period between 2:00AM and 7:30AM (see
also~\autoref{fig:access}), although longer periods (where users are offline for
at least a day) are also regularly observed in the dataset. To mitigate this, we
propose the following feature: if a user's view has \emph{not} been updated for
over \emph{eight hours}, then an alert is triggered. This is incorporated in the
computation of the AADT by removing all time periods that correspond to at least
eight hours of inactivity. 

\subsubsection*{Stratified Sampling}

In the following, we will also divide the servers into several \emph{tiers}:
tier 1 contains servers with 1601-3200 unique users access, tier 2 those with
801-1600 access, up to tier 6 with less than 100 access. If we choose a random
sample $S$ of servers, the tiers that the chosen servers belong have a strong
impact on our results --- e.g., if all the servers are drawn from tiers 5-6 then
the coverage will be poor, whereas even one server from tier 1 would improve the
results dramatically. Hence, in~\autoref{sec:results} we will use a stratified
sampling method that ensures that our sample contains servers from each of tiers
1-4.

\subsubsection{Results} \label{sec:results}

The tables for the empirical results for the coverage, AADT, and average server
freshness in Tables~\autoref{tab:coverage},~\autoref{tab:aadt},
and~\autoref{tab:freshness} respectively. In all of our experiments, we group
the servers in our dataset into five groups such that servers in each group
either belong to the same organization or within a specific IP range. This is
only a way of selecting the servers to run our protocol to show the
effectiveness, there is no strict rule on which servers can adopt our protocol.
Any public server offering a service to the users can help with gossiping, more
the number of servers the faster the attack detection.  

\begin{table}[!h]
\centering
\small
\begin{tabular}{ccccc} 
 & \multicolumn{2}{c}{tier 1} & \multicolumn{2}{c}{tiers 1-4} \\  
  Groups & $n_s$ & Coverage & $n_s$ & Coverage \\ 
  \midrule
  Group 1 & 107 & 0.996 & 731 & 0.996 \\
  Group 2 & 23 & 0.99 & 399 & 0.991 \\
  Group 3 & 32 & 0.981 & 1642 & 0.99 \\
  Group 4 & 9 & 0.982 & 42 & 0.982 \\
  Group 5 & 10 & 0.93 & 64 & 0.953 \\
  \midrule
\end{tabular}
\caption{Results for the coverage (in hours) for the servers of different five different groups.}
\label{tab:coverage}
\end{table}

\noindent In \autoref{tab:coverage}, we display the coverage results for the
five groups.

\begin{table}[!h]
\centering
\small
\begin{tabular}{ccccc} 
 & \multicolumn{2}{c}{tier 1} & \multicolumn{2}{c}{tiers 1-4} \\  
  Groups & $n_s$ & AADT & $n_s$ & AADT \\ 
  \midrule
  Group 1 & 107 & 0.793 & 731 & 0.788 \\
  Group 2 & 23 & 0.862 & 399 & 0.843 \\
  Group 3 & 32 & 1.011 & 1642 & 0.868 \\
  Group 4 & 9 & 0.944 & 42 & 0.929 \\
  Group 5 & 10 & 1.122 & 64 & 1.13 \\
  \midrule
\end{tabular}
\caption{Results for the AADT (in hours) for the servers to five different groups.}
\label{tab:aadt}
\end{table}

 In~\autoref{tab:aadt}, we display the results for the AADT during the active
 periods, averaged across all 2511 users. We see that in each case, the average
 time to detect is around one hour. This can be achieved by just a handful of
 popular servers, e.g., Group 4's 9 or Group 5's 10 most popular servers. In
 most cases, the lower-tier servers do not contribute much to performance ---
 the exception is Group 3, which has a very large number of lower-tier servers,
 which is probably due the servers being part of a large cloud service provider.

\begin{table}[!h]
\centering
\small
\begin{tabular}{ccrclrcl} 
  $p_u$ & $n_u$ & \multicolumn{3}{c}{tier 1} \hspace{-0.42cm} & \multicolumn{3}{c}{tiers 1-4} \hspace{-0.42cm} \\ \midrule
  0.01 & 25 & 0.801 & \hspace{-0.42cm} $\pm$ \hspace{-0.42cm} & 0.0619 & 1.716 & \hspace{-0.42cm} $\pm$ \hspace{-0.42cm} & 0.0566 \\
  0.03 & 75 & 0.492 & \hspace{-0.42cm} $\pm$ \hspace{-0.42cm} & 0.0478 & 1.553 & \hspace{-0.42cm} $\pm$ \hspace{-0.42cm} & 0.0426  \\
  0.1 & 251 & 0.264 & \hspace{-0.42cm} $\pm$ \hspace{-0.42cm} & 0.017 & 1.137 & \hspace{-0.42cm} $\pm$ \hspace{-0.42cm} & 0.0282  \\
  0.3 & 753 & 0.155 & \hspace{-0.42cm} $\pm$ \hspace{-0.42cm} & 0.0045 & 0.863 & \hspace{-0.42cm} $\pm$ \hspace{-0.42cm} & 0.014 \\
  1.0 & 2511 & 0.109 & \hspace{-0.42cm} $\pm$ \hspace{-0.42cm} & 0.0 & 0.591 & \hspace{-0.42cm} $\pm$ \hspace{-0.42cm} & 0.0  \\
  \midrule
\end{tabular}
\caption{95\%-confidence intervals for the average server freshness (in hours) for different user adoption percentages $p_u$. We use random sampling with 8 experiments.}
\label{tab:freshness}
%\vspace{-4mm}
\end{table}

\noindent In~\autoref{tab:freshness}, we have displayed the results for the
server freshness. We considered different user adoption rates, ranging from 1\%
to 100\%. For each of the given percentages, we draw a random sample among the
users, and repeat this experiment 8 times to create  95\% confidence interval
for each entry. We see that even 25 active users can maintain an average server
freshness of below one hour. Note that by a 100\% adoption rate we just refer to
the users in this dataset, and that for a real highest-tier Group 4 \& 5
(servers in this group mainly offers social networking services.) servers, 2500
active lightweight client users may not be unrealistic.

\subsection{The Active Gossip-Based Protocol}
\label{sec:active}
We have seen that the passive mode gossip protocol reduces average attack times
from three hours to around one hour. To improve this even further, the
\emph{active mode} gossip protocol initiates connections to known
protocol-running servers before certain events. One typical event is a user
checking her wallet balance using the client -- since, she is at the risk of
making erroneous decisions  if her client is being eclipsed as part of a
double-spend attack. To obtain knowledge of protocol-running servers, the client
builds a list of servers while it does passive gossiping. That is, if the server
is able to interpret the additional \textit{range} field in the HTTP(S) request
and send back the requested block headers within that range in the corresponding
HTTP(S) response, then client in addition to running our protocol (passive) add
the IP address of the server to the list, if previously unknown. After a
triggering event, a random selection of the known servers are polled from this
list -- the size of this selection depends on the client, as more servers means
a better chance of detection, but more additional traffic. If an up-to-date
server is contacted, then eclipse attack detection is almost instantaneous
during periods of interest.   

In some cases, e.g., if the server is hosting a website associated with a block
explorer (e.g., {\tt blockchain.com}) or a cryptocurrency exchange (e.g., {\tt
binance.com}), then the company that runs the server is likely to run its own
full node anyway, which makes a link even more natural.

\section{Discussion}
\label{sec:discussion}
 In this section, we discuss the real-world implementation of our approach. As
 the deployment of the timestamp-based approach is straightforward (see
 \autoref{sec:baseline}), we limit this discussion to the gossip protocol.

\subsection{Real-World Deployment}
In contrast to competing techniques~\cite{apostolaki2018sabre}, one of our
design goals is to not require any additional infrastructure.  The gossip protocol
works in a distributed way, as anyone can join the network by launching a
supporting server. The dataset used for the analysis mainly consist of traffic
towards web service providers whom we assumed to be willing to deploy our gossip
protocol on their servers. We then evaluated the time required for a typical
Internet user to detect an eclipse attack (analyzed based on our network traffic
data). We selected the most popular (by user interactions) Internet servers to
model the typical pattern of connections between a user and a service. However,
in practice, it is difficult to determine which servers or organizations would
be willing to support the protocol, and we see this as a limitation of our
analysis. 

Finally, we note that although we considered adoption by specific set of servers
categorized into five groups for our analysis, we do not introduce any
centralization factors with it since any number of different entities can run
our protocol simultaneously and independently.

\subsection{Privacy} 

Our protocol is privacy-friendly since lightweight clients gossip only block
headers and do not execute any payment verification with the gossip servers.
Hence, neither Bitcoin wallet addresses nor transaction information is conveyed
to the server. Although lightweight clients reveal to the server their public IP
and the fact that they use Bitcoin, it is not straightforward for an adversarial
server to map the IP to any specific Bitcoin address or transaction. It would be
possible to do so by running Sybil nodes in the Bitcoin network
itself~\cite{chainalysis2015, biryukov2014deanonymisation,biryukov2019security},
but such a threat is orthogonal to our protocol.

\section{Implementation and Evaluation}
\label{sec:implementation}
We implemented our gossip-based protocol using the Python-Flask web
framework~\cite{flask}. The client side uses the Python Requests library that enables it
to send custom HTTP request to servers with modified HTTP headers. We use
Flask's default web server to run the web application that supports our
gossip-based protocol. At the server, block headers are buffered in its memory for
faster access -- however, we leave performance optimization as future work. In
order to send the block headers which are stored as bytes, we use Base64 encoding to
convert the bytes to strings before appending them to the HTTP header.

\subsection*{Evaluation}
\label{subsec:Evaluation}

We have built a custom testbed on Amazon AWS and then used it to evaluate our
gossip-layer implementation. We conducted a series of experiments, picking a
pair of virtual machine instances (t2.micro) running Linux Ubuntu 18.04 in two
different cities across different continents, such that one acts as a web server and the
other as a client. 

\setlength{\tabcolsep}{7pt}
\begin{table}[ht]
\small
\begin{tabular}{cc|cccc}
    &\multicolumn{5}{c}{\textbf{Servers}}\\
    \multirow{10}{*}{\rotatebox{90}{\textbf{Clients}}}
    &&OH&SG&FR&SY\\
    \cline{3-6}
    &\multirow{2}{*}{OH}&\cellcolor{ black!30 }&$0.15\%$&$0.12\%$&$0.12\%$\\
    &&\cellcolor{ black!30 }&$(1.13ms)$&$(0.49ms)$&$(0.86ms)$\\
    &\multirow{2}{*}{SG}&$0.11\%$&\cellcolor{ black!30 }&$0.11\%$&$0.04\%$\\
    &&$(0.96ms)$&\cellcolor{ black!30 }&$(0.69ms)$&$(0.26ms)$\\
    &\multirow{2}{*}{FR}&$0.28\%$&$0.05\%$&\cellcolor{ black!30 }&$0.06\%$\\
    &&$(1.06ms)$&$(0.29ms)$&\cellcolor{ black!30 }&$(0.65ms)$\\
    &\multirow{2}{*}{SY}&$0.09\%$&$0.23\%$&$0.11\%$&\cellcolor{ black!30 }\\
    &&$(0.72ms)$&$(1.66ms)$&$(1.22ms)$&\cellcolor{ black!30 }\\
\end{tabular}
\caption{The average percentage latency inflation with our enhancement when 72 block headers (12 hrs) are transferred.}
\label{tab:httpheader_12}
\end{table}

To evaluate the latency inflation introduced by gossip messages, we measured the
time for a HTTP request-response to complete in two different scenarios, case 1)
a normal HTTP request-response, case 2) A HTTP request-response with an
additional payloads of 72 block headers (12 hrs) in the HTTP header. The time
taken for each case was measured and repeated for 100 such HTTP
request-response. The difference in time between the two cases is the introduced
latency measured. The experiment was redone across different cities. The cities
chosen are Ohio (OH), Singapore (SG), Frankfurt (FR), and Sydney
(SY).~\autoref{tab:httpheader_12} presents the average latency overhead while
the client gossip 72 (12 hrs) of its block headers. The row and column are the
locations of the client and the server respectively. The results show that the
average latency inflation is just 0.12\% which is negligible.

\section{Conclusions}
\label{sec:conclusions}
In this work, we have presented two lightweight protocols for the detection of eclipse
attacks on Bitcoin clients.  Our schemes either use block timestamps, or existing
web servers to create an out-of-Bitcoin gossiping network basing on natural
Internet traffic. We do not require any changes to the Bitcoin protocol or its
network. Moreover, we demonstrate the effectiveness of our protocol by
conducting a series of simulations using real Internet client-server
communication traces and show that it is very efficient. We implemented the
scheme by extending a web application to gossip block headers in its HTTP(S)
header. We proposed a method for compressing Bitcoin subchains and evaluated our
implementation with multiple experiments. Our analysis and evaluation indicate
that the protocol provides multiple benefits, while introducing low overheads
and a significantly improved detection time. Although our system was designed
for Bitcoin, it can be directly applicable to Bitcoin forks and similar
strongest-chain proof-of-work blockchains. Furthermore, we envision that the
protocol can be extended to other classes of blockchain systems, like
proof-of-stake schemes or permissioned blockchains, however we leave this as
future work.

\bibliographystyle{abbrv}
\bibliography{ref}

\appendices
\section{TCP Gossip Layer}
\label{appendix:gossiptcp}

Another design choice of our architecture is implement the message exchange of
our protocol as an extension of the transport layer, more specifically the TCP
protocol. TCP is the de facto standard transport protocol of the Internet,
ubiquitously used and implemented by a large number of existing protocols,
devices, and operating systems. Therefore, any system running a modified TCP
stack with the ability to store and process gossip messages is able to use the
proposed protocol. Another advantage of using TCP as a gossip layer is its
requires no changes to existing applications unlike our present implementation
with HTTP. For these reasons, the adoption of our protocol could be accelerated
and our protocol could perform better in eclipse attack detection (as the
protocol can be executed seamlessly between a large number of clients and
servers). 

\begin{figure}[t!]
    \def\svgwidth{\columnwidth}
    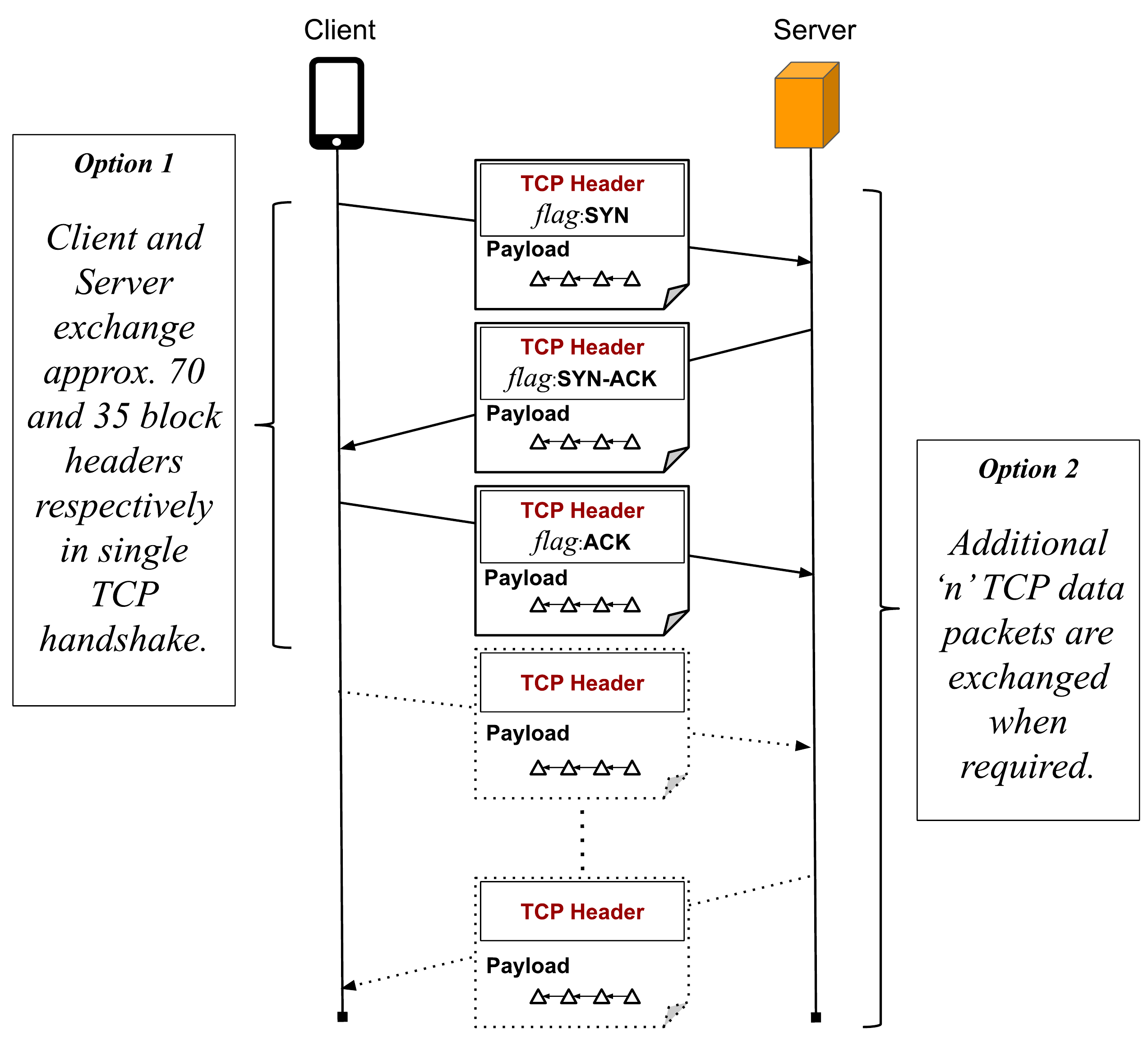   
    \caption{Our protocol implemented by extending a 3-way-TCP handshake between
    a client and server. Option 1 is used when all data requests fits in a
    3-Way-Handshake. Option 2 is used to exchange additional data packets.}     
    \label{fig:gossip}
\end{figure}

More concretely, to minimize the latency inflation caused by sending Bitcoin
headers, we implement our message exchange as data piggybacked on the TCP
handshake. We leverage the fact that the TCP handshake consists of three
messages that do not carry data.\footnotemark Hence, we minimize introduced
latency by exchanging block headers via TCP handshake messages (sent anyway
while establishing a TCP connection).
\footnotetext{Carrying data on TCP handshake packets is allowed by the TCP specification, although such data is not passed to applications. It does not influence our protocol as it is implemented at the TCP (not application) layer.}

As seen in \autoref{fig:gossip}, a client initiates a TCP handshake with a
server and two options are supported. In option 1, the requested block headers
are fully exchanged within a 3-way TCP handshake. However, limitations to the
number of block headers that can fit into a TCP packet implies that further
communication may be needed to send any outstanding block headers. In this case,
our protocol introduces latency inflation and this scenario is illustrated as
option 2. The client and server are seen to have additional block headers
delivered in the data exchange that follows. Usually, each TCP packet is limited
to 1500 bytes.\footnote{1500 bytes is the standard Internet MTU~\cite{rfc894}.}
This approximately corresponds to 35 reduced-size block headers (each 40 bytes)
that can be communicated in a single packet of the TCP handshake. Therefore,
with a TCP handshake only (i.e., $SYN$, $SYN$-$ACK$, and $ACK$), the server can
receive up to 70 headers from the client and the client can receive up to 35
headers from the server, which is near to 12 and 6 hours of block header data
respectively (with an average 10 minute inter-block delay).

\subsection*{Implementation}

To implement our protocol we extended an existing TCP stack. The gossip layer is
implemented on Lightweight IP (LwIP)~\cite{dunkels2001design}, a
state-of-the-art userspace TCP implementation used in production by many
systems. With our implementation, the server and client run a modified TCP stack
to communicate on the network through a TAP/TUN interface~\cite{tuntapdevice}.
The  TCP three-way handshake is initiated by calling \texttt{tcp\_connect()}
which in turn calls \texttt{tcp\_enqueue\_flags()}, to build the packets with
the required TCP flags. We modified this API to allow piggybacking our gossiping
messages on TCP handshake packets. The following code snippet gives an overview
of the modified \texttt{tcp\_enqueue\_flags()} function that caters for gossip
messaging.

\begin{verbatim}
tcp_enqueue_flags(pcb, flags, 
                        gossipMsg){
    gossip_len = sizeof(gossipMsg)
    packet = pbuf_alloc(PBUF_TRANSPORT, 
                optlen + gossip_len, 
                PBUF_RAM))
    TCP_DATA_COPY2(packet->payload + 
                    optlen, gossipMsg)
    seg = tcp_create_segment(pcb, 
                    packet, flags, 
                    pcb->snd_lbb, 
                    optflags))   
}
\end{verbatim}

%pcb->unsent = seg

The \texttt{tcp\_enqueue\_flags()} function takes three arguments, namely:
protocol control block (pcb), TCP flags, and a gossip message of our protocol.
The \texttt{pbuf\_alloc()} function is responsible for memory allocation and in
our case we have to adjust the parameters of this functions such that an
additional memory for the gossip message is allocated. The
\texttt{TCP\_DATA\_COPY2()} function is a variant of a memory copy function used
to copy the gossip message onto the packet. The \texttt{tcp\_create\_segment()}
function creates a TCP segment, with the required gossip message. At the
receiving end, TCP segments are handled by the \texttt{tcp\_input()} function.

\begin{verbatim}
tcp_input(tcp_segment){
    removeHeaders(tcp_segment->payload)
    MEMCPY(buffer, tcp_segment->payload, 
                    tcp_segment->length)
}
\end{verbatim}

With this call, the packet header in \texttt{tcp\_segment->payload} is removed
and the received gossip message (block headers) is left. It is copied to the
\texttt{buffer} for further processing of our protocol.  We implemented a simple
file transfer (client-server) application using our modified stack.

\subsection*{Evaluation}

The setup is similar to~\autoref{subsec:Evaluation},

\setlength{\tabcolsep}{7pt}
\begin{table}[ht]
\small
\begin{tabular}{cc|cccc}
    &\multicolumn{5}{c}{\textbf{Servers}}\\
    \multirow{10}{*}{\rotatebox{90}{\textbf{Clients}}}
    &&OH&SG&FR&SY\\
    \cline{3-6}
    &\multirow{2}{*}{OH}&\cellcolor{ black!30 }&$2.26\%$&$2.39\%$&$1.22\%$\\
    &&\cellcolor{ black!30 }&$(2.22ms)$&$(2.33ms)$&$(2.35ms)$\\
    &\multirow{2}{*}{SG}&$1.09\%$&\cellcolor{ black!30 }&$1.47\%$&$1.32\%$\\
    &&$(2.39ms)$&\cellcolor{ black!30 }&$(2.52ms)$&$(2.31ms)$\\
    &\multirow{2}{*}{FR}&$2.36\%$&$1.28\%$&\cellcolor{ black!30 }&\multirow{2}{*}{0}\\
    &&$(2.30ms)$&$(2.22ms)$&\cellcolor{ black!30 }&\\
    &\multirow{2}{*}{SY}&$1.23\%$&$1.32\%$&\multirow{2}{*}{0}&\cellcolor{ black!30 }\\
    &&$(2.37ms)$&$(2.22ms)$&&\cellcolor{ black!30 }\\
\end{tabular}
\caption{The average percentage latency inflation with our enhancement.}
\label{tab:tcphandshake}
\end{table}

In this case also, we measured the time for a TCP connection establishment in
two different scenarios, case 1) A normal TCP handshake, case 2) A TCP handshake
with an additional payload of 1440 bytes (including 35 block headers). The time
taken for each case was measured and repeated for 100 such TCP handshakes. The
difference in time between the two cases is the introduced latency measured. The
experiment was redone across different cities.~\autoref{tab:tcphandshake}
presents the average latency overhead. On an average, the overhead is seen to
be close to $1.59\%$. For handshakes between Frankfurt and Sydney, no overhead
was recorded and may be attributed to high network latency.

\section{HTTP body Gossip-Layer}
\label{appendix:http}
In this section we present the results for evaluation of our gossiping using
HTTP body. Unlike sending data in HTTP headers of request-response message.
Here, we provide an implementation based on sending the block headers as file
chunks in HTTP body with 72 (12 hrs), 144 (24 hrs), 1008 (7 days) and 2016 (14
days) block headers. The results are presented in
tables~\autoref{tab:httpconnection_12},~\autoref{tab:httpconnection_24},~\autoref{tab:httpconnection_14d},
and~\autoref{tab:httpconnection_7d} respectively. The experimental setup and
evaluation methodology is exactly same as in~\autoref{subsec:Evaluation}.

\setlength{\tabcolsep}{2.5pt}
\begin{table}[ht]
\small
\centering
\begin{tabular}{cc|cccc}
    &\multicolumn{5}{c}{\textbf{Servers}}\\
    \multirow{10}{*}{\rotatebox{90}{\textbf{Clients}}}
    &&OH&SG&FR&SY\\
    \cline{3-6}
    &\multirow{2}{*}{OH}&\cellcolor{ black!30 }&$149.56\%$&$148.77\%$&$152.82\%$\\
    &&\cellcolor{ black!30 }&$(1338.74ms)$&$(1162.44ms)$&$(579.62ms)$\\
    &\multirow{2}{*}{SG}&$149.94\%$&\cellcolor{ black!30 }&$143.03\%$&$153.66\%$\\
    &&$(1341.85ms)$&\cellcolor{ black!30 }&$(982.34ms)$&$(1106.01ms)$\\
    &\multirow{2}{*}{FR}&$149.09\%$&$163.08\%$&\cellcolor{ black!30 }&$152.02\%$\\
    &&$(581.06s)$&$(1094.42ms)$&\cellcolor{ black!30 }&$(1740.01ms)$\\
    &\multirow{2}{*}{SY}&$150.14\%$&$176.30\%$&$150.78\%$&\cellcolor{ black!30 }\\
    &&$(1142.51ms)$&$(1269.78ms)$&$(1724.61ms)$&\cellcolor{ black!30 }\\
\end{tabular}
\caption{The average percentage latency inflation with our enhancement when 2016 block headers (14 days) are transferred.}
\label{tab:httpconnection_14d}
\end{table}

\setlength{\tabcolsep}{4pt}
\begin{table}[ht]
\small
\begin{tabular}{cc|cccc}
    &\multicolumn{5}{c}{\textbf{Servers}}\\
    \multirow{10}{*}{\rotatebox{90}{\textbf{Clients}}}
    &&OH&SG&FR&SY\\
    \cline{3-6}
    &\multirow{2}{*}{OH}&\cellcolor{ black!30 }&$99.63\%$&$99.32\%$&$128.35\%$\\
    &&\cellcolor{ black!30 }&$(891.71ms)$&$(386.94ms)$&$(976.29ms)$\\
    &\multirow{2}{*}{SG}&$99.58\%$&\cellcolor{ black!30 }&$99.02\%$&$100.68\%$\\
    &&$(0.19ms)$&\cellcolor{ black!30 }&$(0.53ms)$&$(1.46ms)$\\
    &\multirow{2}{*}{FR}&$99.31\%$&$100.72\%$&\cellcolor{ black!30 }&$99.87\%$\\
    &&$(387.01ms)$&$(675.87ms)$&\cellcolor{ black!30 }&$(1143.12ms)$\\
    &\multirow{2}{*}{SY}&$99.57\%$&$100.38\%$&$99.76\%$&\cellcolor{ black!30 }\\
    &&$(757.62ms)$&$(722.98ms)$&$(1141.07ms)$&\cellcolor{ black!30 }\\
\end{tabular}
\caption{The average percentage latency inflation with our enhancement when 1008 block headers (7 days) are transferred.}
\label{tab:httpconnection_7d}
\end{table}

\setlength{\tabcolsep}{7pt}
\begin{table}[ht]
\small
\begin{tabular}{cc|cccc}
    &\multicolumn{5}{c}{\textbf{Servers}}\\
    \multirow{10}{*}{\rotatebox{90}{\textbf{Clients}}}
    &&OH&SG&FR&SY\\
    \cline{3-6}
    &\multirow{2}{*}{OH}&\cellcolor{ black!30 }&$0.04\%$&$0.21\%$&$0.03\%$\\
    &&\cellcolor{ black!30 }&$(0.27ms)$&$(0.79ms)$&$(0.21ms)$\\
    &\multirow{2}{*}{SG}&$0.10\%$&\cellcolor{ black!30 }&$0.04\%$&$0.06\%$\\
    &&$(0.89ms)$&\cellcolor{ black!30 }&$(0.41ms)$&$(0.41ms)$\\
    &\multirow{2}{*}{FR}&$0.20\%$&$0.02\%$&\cellcolor{ black!30 }&$0.04\%$\\
    &&$(0.80ms)$&$(0.11ms)$&\cellcolor{ black!30 }&$(0.42ms)$\\
    &\multirow{2}{*}{SY}&$0.01\%$&$0.13\%$&$0.07\%$&\cellcolor{ black!30 }\\
    &&$(0.05ms)$&$(0.9ms)$&$(0.69ms)$&\cellcolor{ black!30 }\\
\end{tabular}
\caption{The average percentage latency inflation with our enhancement when 72 block headers (12 hrs) are transferred in HTTP body.}
\label{tab:httpconnection_12}
\end{table}

\setlength{\tabcolsep}{7pt}
\begin{table}[ht]
\small
\begin{tabular}{cc|cccc}
    &\multicolumn{5}{c}{\textbf{Servers}}\\
    \multirow{10}{*}{\rotatebox{90}{\textbf{Clients}}}
    &&OH&SG&FR&SY\\
    \cline{3-6}
    &\multirow{2}{*}{OH}&\cellcolor{ black!30 }&$0.09\%$&$0.25\%$&$0.07\%$\\
    &&\cellcolor{ black!30 }&$(0.78ms)$&$(1.01ms)$&$(0.48ms)$\\
    &\multirow{2}{*}{SG}&$0.21\%$&\cellcolor{ black!30 }&$0.16\%$&$0.13\%$\\
    &&$(1.92ms)$&\cellcolor{ black!30 }&$(1.10ms)$&$(0.91ms)$\\
    &\multirow{2}{*}{FR}&$0.28\%$&$0.28\%$&\cellcolor{ black!30 }&$0.07\%$\\
    &&$(1.08ms)$&$(1.85ms)$&\cellcolor{ black!30 }&$(0.7ms)$\\
    &\multirow{2}{*}{SY}&$0.04\%$&$0.32\%$&$0.08\%$&\cellcolor{ black!30 }\\
    &&$(0.30ms)$&$(2.29ms)$&$(0.84ms)$&\cellcolor{ black!30 }\\
\end{tabular}
\caption{The average percentage latency inflation with our enhancement when 144 block headers (24 hrs) are transferred.}
\label{tab:httpconnection_24}
\end{table}

\begin{figure}[t!]
  \begin{subfigure}{.48\textwidth}
    \centering
    \def\svgwidth{\columnwidth}
    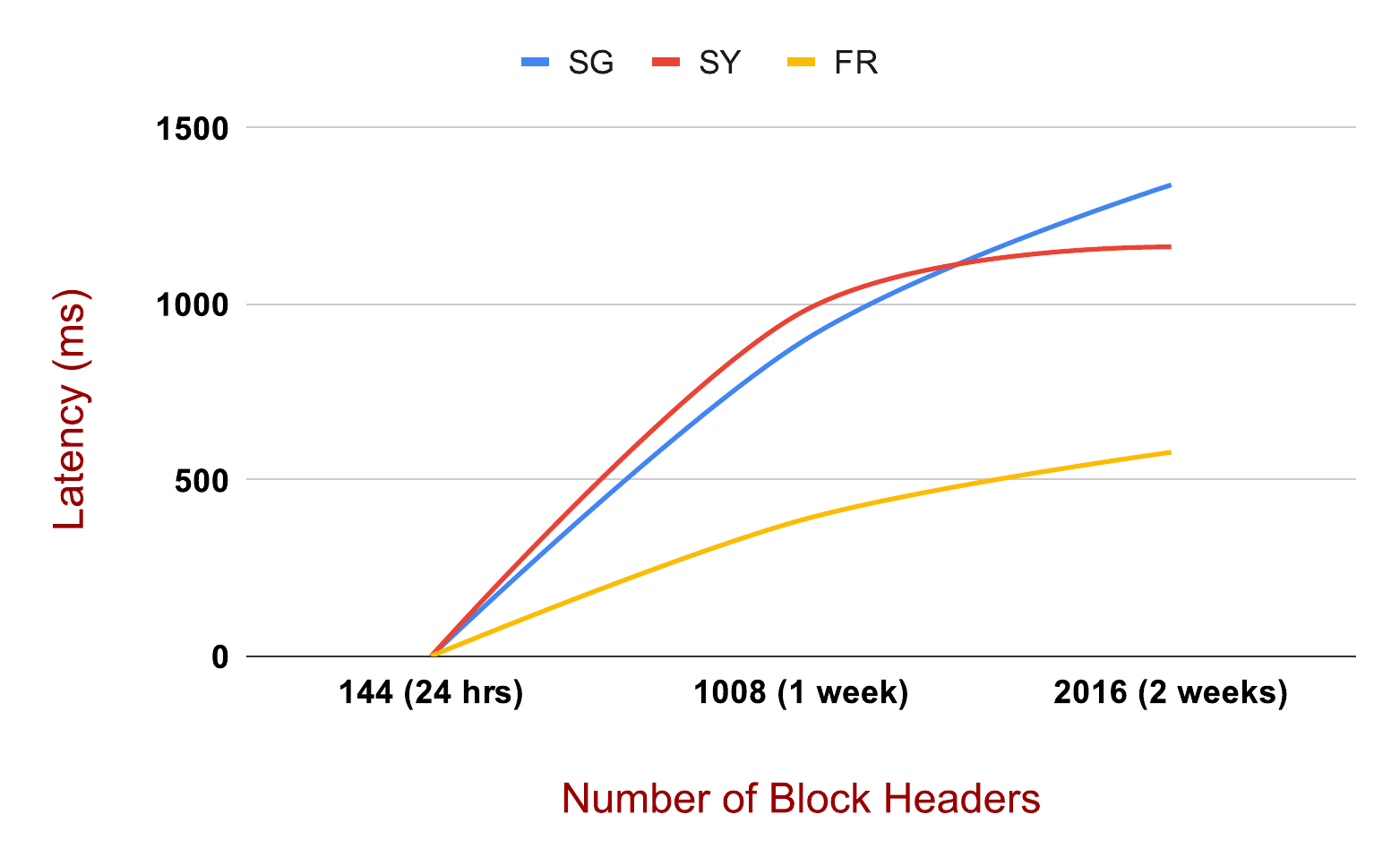   
    \caption{Client OH}     
    \label{fig:client_oh}
  \end{subfigure}
  \begin{subfigure}{.48\textwidth}
    \centering
    \def\svgwidth{\columnwidth}
    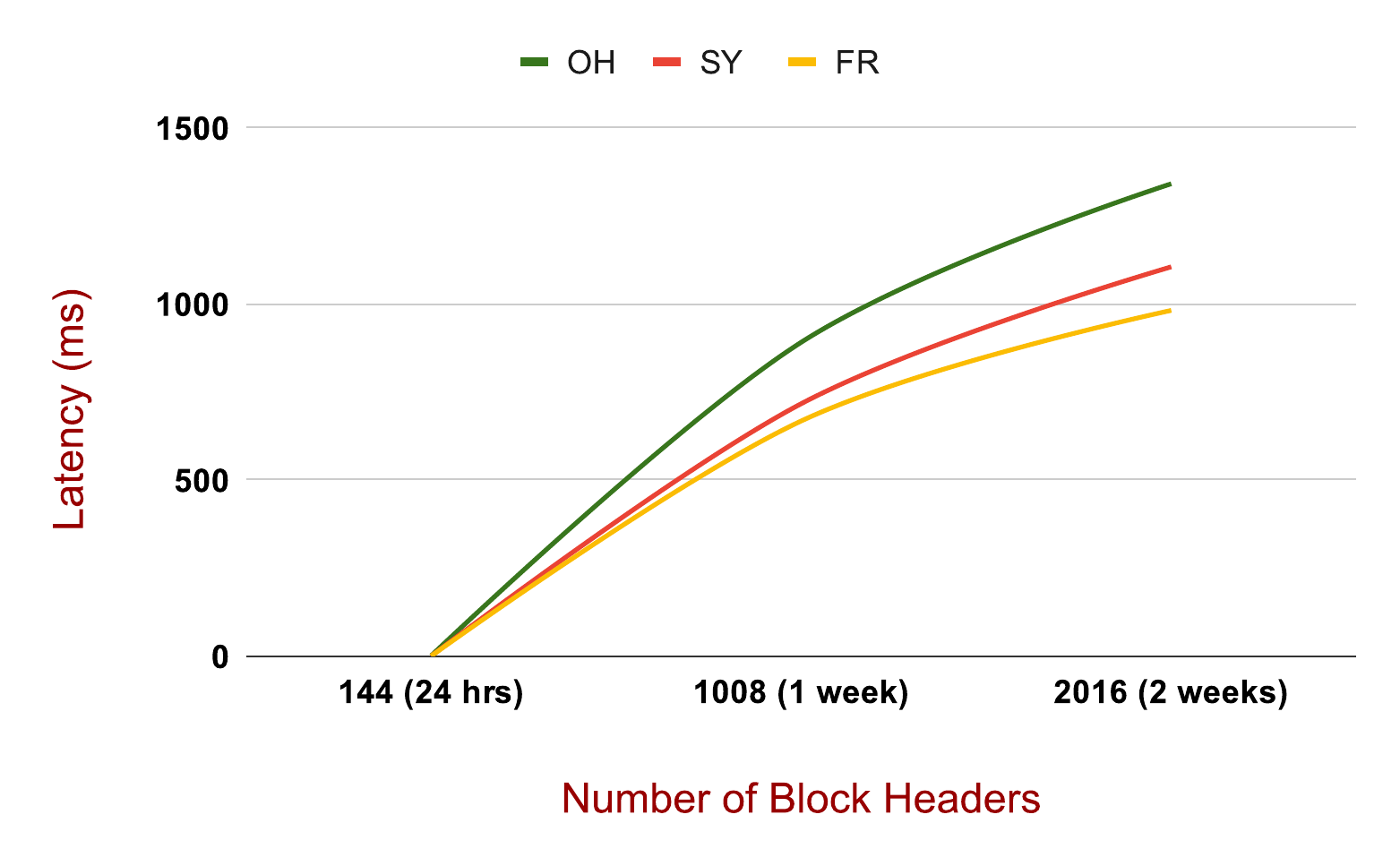   
    \caption{Client SG}     
    \label{fig:client_sg}
  \end{subfigure}
  \caption{(a) The average latency when client is at OH and gossips with server at
  SG, SY and FR, (b) The average latency when client is at SG and gossips with
  server at OH, SY and FR.}
  \label{fig:client_latency}
  \end{figure}
  
  \begin{figure}[t!]
    \vspace{-50mm}
  \begin{subfigure}{.48\textwidth}
    \centering
  
    \centering
    \def\svgwidth{\columnwidth}
   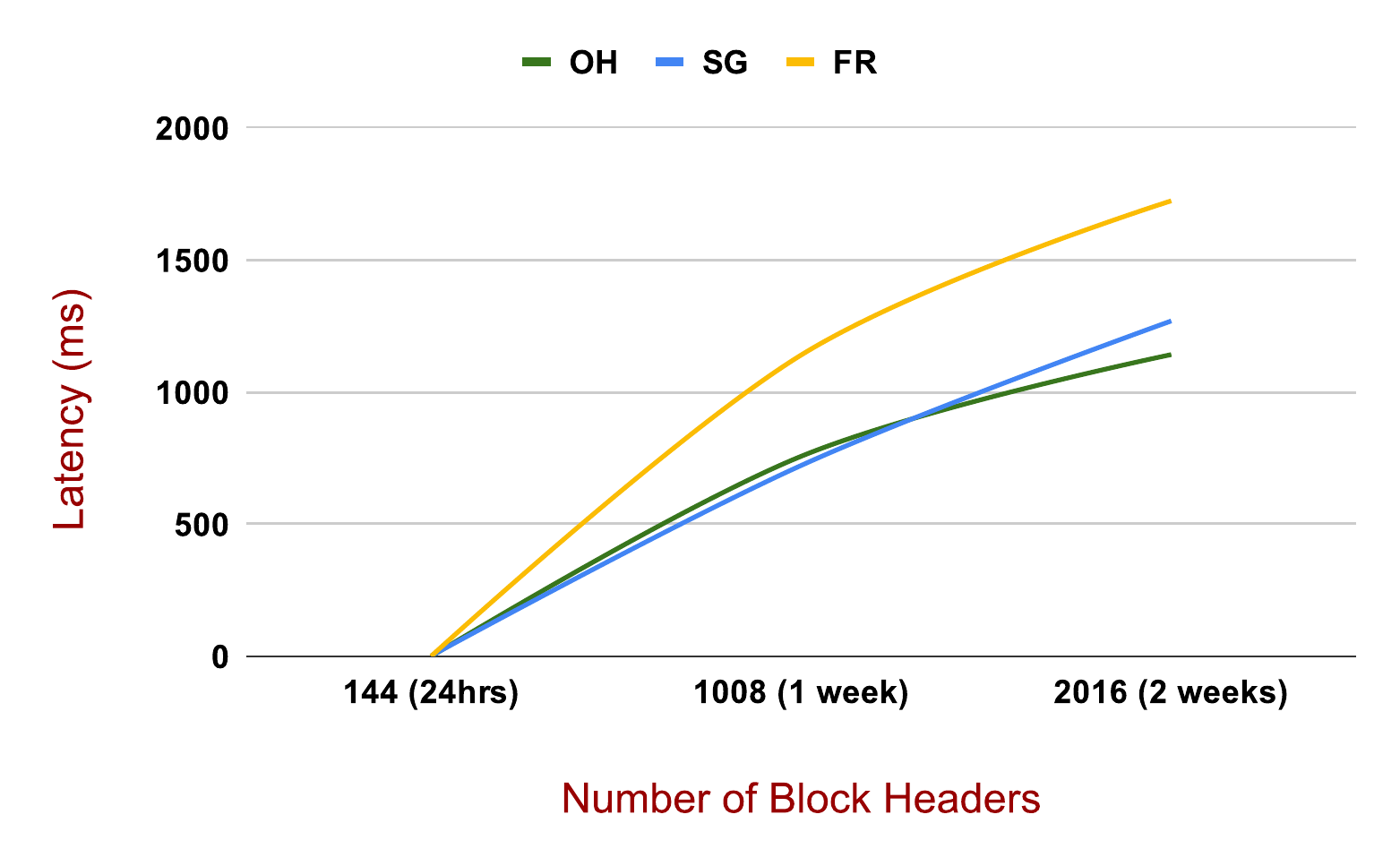   
    \caption{Client SY} 
    \label{fig:client_sy}    
   \end{subfigure}
  \begin{subfigure}{.48\textwidth}
    \centering
    \def\svgwidth{\columnwidth}
    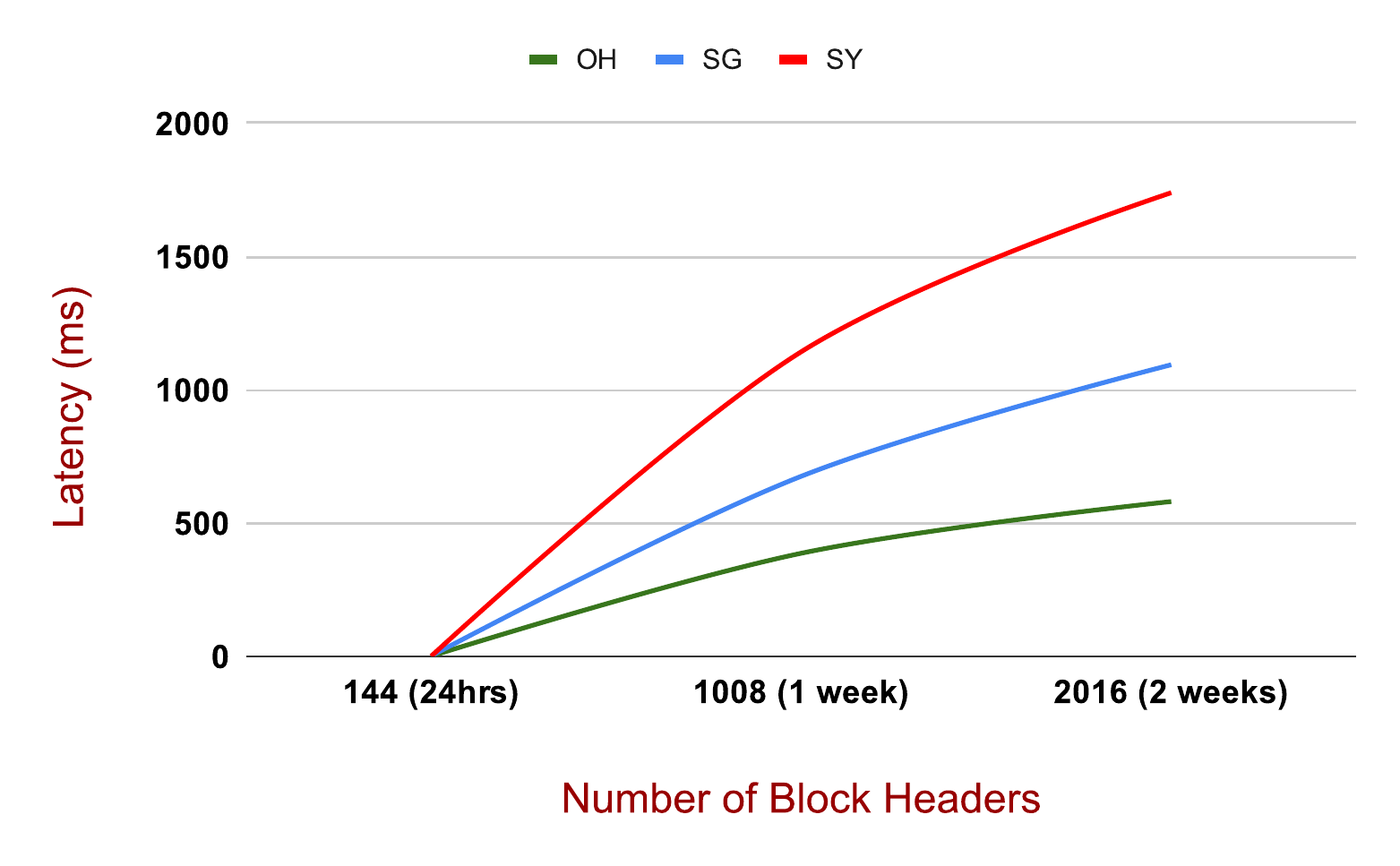   
    \caption{Client FR}     
    \label{fig:client_fr}
    \end{subfigure}
    \caption{(a) The average latency when client is placed at SY and gossip with
  server at OH, SG and FR, (b) The average latency when client is placed at SG and
  gossip with server at OH, SY and FR.}
  \label{fig:client_fr_sy}
  \end{figure}

As we can see the latency inflation is almost on an average 100\% for block
headers generated over a week and increases by 50-60\% for two weeks. In case of
12 hrs i.e. 72 block headers the average latency inflation is 0.08\% and
increase with 24 hrs block headers. The results indicate that the inflation
remains negligible if the users come online for at least once in a day which is
not unrealistic with the present day Internet users.  

In~\autoref{fig:client_latency} and~\autoref{fig:client_fr_sy} show the average
latency for HTTP request-response for clients with our enhance while they are
placed at OH, SG, FR and SY. The results prove that the latency only grows
linearly with increase in number of block headers.

\end{document}